\documentclass[a4paper,11pt]{article}
\usepackage[utf8]{inputenc}
\usepackage[english]{babel}
\usepackage[T1]{fontenc}
\usepackage[margin=2.5cm]{geometry}
\usepackage{amsmath}
\usepackage{hyperref}
\usepackage[numbers]{natbib}
\usepackage{amssymb}
\usepackage{graphicx}
\usepackage{caption}
\usepackage{subcaption}
\newcounter{subfig}

\usepackage{booktabs}
\usepackage{tikz}
\usepackage{braket}
\usepackage{xcolor}
\newtheorem{theorem}{Theorem}
\newtheorem{lemma}{Lemma}

\begin{document}

\title{Hybrid Gaussian-exponential zero-noise extrapolation for periodic circuits}

\author{Tao Wang$^{1,2}$ and Yun Shang$^{1,3,\ast}$}
\date{}

\maketitle

\noindent
$^{1}$Institute of Mathematics, Academy of Mathematics and Systems Science,\\
Chinese Academy of Sciences, Beijing 100190, China\\
$^{2}$School of Mathematical Sciences, University of Chinese Academy of Sciences,\\
Chinese Academy of Sciences, Beijing 100049, China\\
$^{3}$State Key Laboratory of Mathematical Sciences, Academy of Mathematics and Systems Science,\\
Chinese Academy of Sciences, Beijing 100190, China\\
$^{\ast}$\texttt{shangyun@amss.ac.cn}

\begin{abstract}
  Zero-noise extrapolation provides a practical means of suppressing gate errors in current noisy intermediate-scale quantum hardware. The accuracy of the zero-noise estimate depends sensitively on the fidelity of the assumed noise model to the actual error scaling. This work introduces a hybrid Gaussian-exponential extrapolation scheme tailored for quantum circuits with periodic structure, which are common in quantum algorithms. Under Pauli diagonal errors, by constructing and analyzing an approximate Markov process for the transfer of Pauli operators, we prove a central limit theorem: the noise amplification factor weakly approaches a log-normal distribution, which motivates augmenting the standard exponential model with Gaussian variance corrections. The resulting model requires no prior noise characterization and applies directly to arbitrary periodic circuits. Performance is assessed on Trotterized Ising dynamics, random circuits, and Grover search algorithm using Qiskit noise simulators. For moderate to large circuit depths, the hybrid model yields clear reductions in bias relative to previous extrapolation variants, showing its usefulness for error mitigation on near-term quantum hardware.
\end{abstract}

\section{Introduction}
\label{sec:level1.0}
Current quantum computing devices are still in the
noisy intermediate-scale quantum (NISQ) era \cite{Preskill2018quantumcomputingin}.
At this stage, quantum processors have reached
scales of hundreds of qubits, however gate errors, readout errors,
decoherence, etc., inevitably lead to various
types of noise, significantly reducing the accuracy
of quantum algorithm outputs. In theory, when the system
noise level is suppressed below the fault-tolerance threshold,
fault-tolerant quantum computation can be achieved through
quantum error correction (QEC) \cite{Knill_1998, doi:10.1137/S0097539799359385, Kitaev_2003}.
Still, the practical implementation of quantum error correction
on large-scale devices remains challenging in the near term
due to its extremely high demands on physical qubit counts,
gate fidelities, and control precision. In the absence of full fault tolerance,
how to improve the accuracy of quantum computation results
in the presence of noise has become an important problem
in current quantum computing research.

To address this challenge, a variety of quantum error mitigation (QEM)
techniques have been proposed in recent years.
The basic idea of these methods is not to completely
eliminate noise, but to systematically process noisy
experimental outputs to infer estimates close to the
ideal noise-free expectation values. Compared with
QEC, QEM typically does not require
a large number of extra logical qubits or complex encoding frameworks,
making it well suited for current NISQ devices.

Several QEM approaches have been developed, including zero-noise extrapolation
(ZNE) \cite{PhysRevLett.119.180509, PhysRevX.7.021050},
probabilistic error cancellation (PEC) \cite{PhysRevLett.119.180509, PhysRevX.8.031027},
randomized compiling (RC) \cite{PhysRevA.94.052325},
subspace expansion \cite{McClean:2019rbh},
and learning-based error mitigation \cite{PRXQuantum.2.040330, Liao_2024}.

Among these, zero-noise extrapolation is widely adopted due to its convenience.
Its core idea is to artificially amplify the noise strength of a circuit,
obtain expectation values at different noise levels,
and then extrapolate the data back to the zero-noise limit by
fitting a noise model to estimate the ideal value.
Temme et al. \cite{PhysRevLett.119.180509} proposed a ZNE method that
scales noise by extending pulse durations on physical devices,
but this approach requires high-precision control of physical
parameters and is not universally applicable to all quantum hardware.
Another approach scales noise at the gate level via
unitary folding \cite{Giurgica_Tiron_2020}, however its theoretical
validity relies on specific noise structures, such as depolarizing noise.

To analyze the effect of unitary folding on circuit noise more carefully,
inverse-circuit ZNE (IC-ZNE) \cite{PhysRevA.110.042625} and
purity-assisted ZNE (pZNE) \cite{Jin_2024} have been proposed.
However, these methods still have theoretical limitations. They all rely on simple approximations to model the noise effect on the circuit. Such approximations gradually break down as the circuit depth increases.
In addition, these methods have only been tested on small-scale circuits.

In this work, starting from the circuit model and using
statistical analysis, we introduce the hybrid Gaussian-exponential zero-noise extrapolation (G-E ZNE) method
in the asymptotic limit for periodic circuits subject to Pauli noise \cite{Nielsen_Chuang_2010}. We extract an adjacency matrix structure
from each period block, which induces an approximate
Markov process for the transfer of Pauli operators in the overall circuit.
From this, we prove a central limit theorem: the distribution of noise amplification factors over different transfer processes weakly approaches a log-normal distribution as the number of periods grows. This clarifies the
statistical behavior of noise along different Pauli paths, giving a more accurate description of noise scaling in deep structured circuits.
Using this statistical regularity, we obtain a better approximate
model and achieve more effective error mitigation. Compared with pZNE and IC-ZNE, we characterize the noise effects more precisely, rather than relying on depolarizing noise or averaging approximations of the noise.
For finite periods, we conduct experiments on Qiskit simulators
using Trotterized dynamics of the
one-dimensional Ising spin chain, random circuits, and Grover circuits.
These circuit classes are diverse, and by increasing the number of periods, we can more intuitively observe the mitigation performance at different depths.
From the experimental results, our method achieves lower bias than standard ZNE, IC-ZNE, and pZNE in terms of accuracy, especially in the regime of larger periods.

The remainder of this paper is organized as follows.
Section~\ref{sec:level2.0} introduces the standard ZNE, IC-ZNE, and pZNE methods.
Section~\ref{sec:level3.0} proves a central limit theorem, thus deriving a hybrid Gaussian-exponential extrapolation model
based on the Pauli expansion of quantum channels and states.
Section~\ref{sec:level4.0} presents experimental comparisons on Qiskit simulators
(FakeQuito and FakeLima) with the methods described in Section~\ref{sec:level2.0},
demonstrating the advantages of our extrapolation model.

\section{Preliminaries}
\label{sec:level2.0}
\subsection{Notation}
\label{sec:level2.1}
A quantum state is represented by a positive semidefinite
Hermitian matrix $\rho$ with $\operatorname{tr}(\rho)=1$.
Its evolution is governed by a quantum channel $\mathcal{E}(\rho)$,
which is a completely positive and trace-preserving (CPTP) linear map.
The expectation value of an observable $O$ is given by
\begin{align}
    \langle O \rangle = \operatorname{tr}\bigl( O \,\mathcal{E}(\rho) \bigr).
\end{align}

\subsection{Standard zero-noise extrapolation}
\label{sec:level2.2}
Unitary folding is a noise-scaling technique that replaces a
unitary circuit (or gate) $U$ by $U(U^{\dagger}U)^{r}$, where
$r$ is a positive integer. This operation alters the effective noise strength
while preserving the ideal logical functionality of the circuit.
In the case of depolarizing noise, let $n$ be the number
of qubits affected by noise and $p$  the noise parameter.
The noisy implementation of $U$ is
\begin{align}
    \mathcal{E}_{U}=\mathcal{E}_{p} \circ \mathcal{U},
\end{align}
with
\begin{align}
    \mathcal{E}_{p}(\rho)=(1-p)\rho + p\frac{\mathbf{I}}{2^{n}}.
\end{align}
Consequently,
\begin{align}
     \mathcal{E}_{U} \bigl( \mathcal{E}_{U^{\dagger}} \mathcal{E}_{U} \bigr)^{r} = \mathcal{E}_{1-(1-p)^{2r+1}} \circ \mathcal{U}.
\end{align}
Let $\lambda=2r+1$ denote the noise amplification factor,
we can interpret the combined noise channel as an
equivalently stronger depolarizing channel.
The expectation value of the noisy circuit is then given by
\begin{align}
    \langle O \rangle(\lambda) &= (1-p)^{\lambda} \langle O \rangle_{\text{ideal}} + \left(1-(1-p)^{\lambda}\right) \frac{\operatorname{tr}(O)}{2^{n}} \nonumber\\
    &= (1-p)^{\lambda} \left( \langle O \rangle_{\text{ideal}} - \frac{\operatorname{tr}(O)}{2^{n}} \right) + \frac{\operatorname{tr}(O)}{2^{n}}.
\end{align}
Defining $a = \langle O \rangle_{\text{ideal}} - \operatorname{tr}(O)/2^{n}$,
$b = 1-p$, and $c = \operatorname{tr}(O)/2^{n}$, we obtain the
exponential extrapolation model
\begin{align}
    \langle O \rangle(\lambda) = a b^{\lambda} + c,
\end{align}
from which the ideal expectation value $\langle O \rangle_{\text{ideal}}$ is
recovered by extrapolating $\lambda \to 0$.

In addition to exponential fitting, widely adopted techniques
such as polynomial fitting and Richardson extrapolation\cite{PhysRevLett.119.180509}
are also employed. Under general Pauli noise, however,
these models may exhibit a non-negligible discrepancy
from the true noise behavior.
Reference~\cite{Cai_2021} generalizes exponential extrapolation
to a multi-exponential form that captures multiple decay modes:
\begin{align}
    \langle O \rangle(\lambda) = \sum_{k=1}^{K} a_k b_k^{\lambda} + c,
\end{align}
where $k$ counts the erroneous gate operations, and
$K$ denotes the maximal number of such errors that can occur.
In the third Grover search experiment of this paper,
we compare our method with the dual-exponential model.

\subsection{Inverted-circuit zero-noise extrapolation}
\label{sec:level2.3}
The IC-ZNE method does not rely on the proportional
relationship between the noise strength and the number
of folding operations. Instead, it indirectly determines
the noise strength by measuring the circuit fidelity.

Let the ideal circuit output be $\ket{\psi}$ and
the noisy circuit output be $\rho$. The fidelity $F$ is defined as
\begin{align}
    F=\braket{\psi|\rho|\psi}.
\end{align}
The noise strength is then defined as
\begin{align}
    \epsilon=1-F.
\end{align}
Consequently, $\rho$ can be decomposed into an operator
$\sigma$ orthogonal to $\ket{\psi}$:
\begin{align}
    \rho=(1-\epsilon)\ket{\psi}\bra{\psi}+\epsilon \sigma.
\end{align}
The expectation value of an observable $O$ is given by
\begin{align}
    \langle O \rangle = (1-\epsilon) \braket{\psi | O | \psi} + \epsilon \operatorname{tr}(O \sigma) \label{eq10}.
\end{align}

The value of $\epsilon$ can be extracted by implementing
the circuit $U^{\dagger}U$ and measuring the probability
of obtaining the outcome $\ket{0}^{\otimes n}$ (the initial state).
This probability is
\begin{align}
    P_{0} &= \langle 0| \bigl( \mathcal{E}_{U^{\dagger}} \mathcal{E}_{U} \bigr)(|0\rangle\langle 0|) |0\rangle \nonumber\\
    &= \operatorname{tr}\!\left\{ \mathcal{E}_{U^{\dagger}}^{\dagger}(|0\rangle\langle 0|) \; \mathcal{E}_{U}(|0\rangle\langle 0|) \right\} \nonumber\\
    &= \operatorname{tr}\{ \tilde{\rho} \, \rho \},
\end{align}
where
\begin{align}
    \tilde{\rho}=\mathcal{E}_{U^{\dagger}}^{\dagger}(|0\rangle\langle 0|),
\end{align}
which is called the dual state\cite{Huo_2022}. And $\mathcal{E}_{U^{\dagger}}^{\dagger}$ is the dual map\cite{Watrous_2018},
satisfying
\begin{align}
    \langle X, \mathcal{E}(\rho) \rangle = \langle \mathcal{E}^\dagger(X), \rho \rangle \quad \forall \rho, X,
\end{align}
where $\langle A, B \rangle = \operatorname{tr}(A^\dagger B)$ is the Hilbert--Schmidt inner product.

The dual state $\tilde{\rho}$ shares the same noise strength as $\rho$\cite{PhysRevA.110.042625}, so
it admits a decomposition
\begin{align}
    \tilde{\rho}=(1-\epsilon)\ket{\psi}\bra{\psi}+\epsilon \tilde{\sigma}.
\end{align}
By construction, $\sigma$ and $\tilde{\sigma}$ have zero overlap with
$\ket{\psi}$, i.e. $\braket{\psi | \sigma | \psi}=\braket{\psi | \tilde{\sigma} | \psi}=0$.
Thus,
\begin{align}
    P_{0}=(1-\epsilon)^{2}+\epsilon^{2}\operatorname{tr}(\sigma \tilde{\sigma}).
\end{align}
Let $a=\operatorname{tr}(\sigma \tilde{\sigma})$. If
$a$ is taken as a constant, it can be treated as a fitting parameter.
In the literature~\cite{PhysRevA.110.042625}, $a$ is set to
$a=1/2^{n}$, which is slightly larger than
its value for depolarizing noise. This leads to the expression
\begin{align}
    \epsilon = \left\{
    \begin{array}{cc}
        \dfrac{1 - \sqrt{P_{0} - \frac{1-P_{0}}{2^{n}}}}{1 + \frac{1}{2^{n}}}, & P_{0} > \dfrac{1}{2^{n}} \\[10pt]
        \dfrac{1 - P_{0}}{1 + P_{0}}, & P_{0} \le \dfrac{1}{2^{n}} .
    \end{array}
    \right.
\end{align}
By extrapolating to $\epsilon \rightarrow{0}$ using a linear fit, we
obtain the estimation of the ideal expectation.

The IC-ZNE method provides a more accurate characterization of
the noise strength.
The disadvantage, however, is that some of its approximations
may not hold universally for general circuits. For example,
parameters such as $\sigma$ and $a$ may no longer remain constant under folding, leading to a greater deviation of the actual model from the ideal one.

\subsection{Purity-assisted zero-noise extrapolation}
\label{sec:level2.4}
The pZNE method similarly does not presuppose a linear
relationship between noise strength and folding factor.
Given a Pauli noise channel $\mathcal{E}$, let
$\lambda_{f,i}$ and $\lambda_{b,i}$ represent its forward and
backward eigenvalues(see Subsection~\ref{sec:level3.1}),
which quantify how the expectation value of each Pauli operator $P_{i}$
is attenuated under the noisy evolution. After folding $n$ times,
\begin{align}
    \langle P_i \rangle_n = \lambda_{f,i}^{\,n+1} \lambda_{b,i}^{\,n} \langle P_i \rangle_{\text{ideal}}.
\end{align}
Define $\chi_{n,i} = \lambda_{f,i}^{\,n+1} \lambda_{b,i}^{\,n}$.
Then the purity of the output state is given by
\begin{align}
    p_n &= \frac{1}{2^{q}} \sum_i \chi_{n,i}^{2} \langle P_i \rangle^2_{\text{ideal}} \nonumber\\
        &= (p_0 - p_\infty) \overline{\chi_n^{2}} + p_\infty,
\end{align}
where $\overline{\chi_n^{2}} = \frac{\sum_{\chi_{n,i}<1} \chi_{n,i}^{2} \langle P_i \rangle^2_{\text{ideal}}}{\sum_{\chi_{n,i}<1} \langle P_i \rangle^2_{\text{ideal}}}$
is the weighted average of the nontrivial eigenvalues.
Provided the variance of the Pauli noise channel's eigenvalues
is sufficiently small, the approximation
\begin{align}
    \chi_{n,i} \approx \overline{\chi_{n}^{2}}^{1/2}=\sqrt{\frac{p_{n}-p_{\infty}}{p_{0}-p_{\infty}}},
\end{align}
holds, leading to the fitting relation between the
noisy expectation value and the purity:
\begin{align}
    \langle P_i \rangle_n = \langle P_i \rangle_{\text{ideal}} \sqrt{\frac{p_n - p_\infty}{p_0 - p_\infty}}.
\end{align}
The error scales with the standard deviation of the noise channel eigenvalues.

A key advantage of pZNE is that it does not rely on a
pre-specified error model. For general Pauli noise, it provides
a rigorous theoretical framework that clarifies why the noise
strength is not a linear function of the folding number and
proposes a purity-based method to mitigate this issue.
Its primary limitation is that the original derivation
is restricted to circuits consisting of individual gates,
and therefore cannot be directly generalized to
arbitrary circuits. In the case of Clifford circuits,
the closure property of Clifford gates on the Pauli group
allows the Pauli noise channel to be regarded as a global entity.
For more general circuits, randomized compiling can be employed to
partition the overall circuit into local segments,
each of which effectively experiences Pauli noise.
Building upon this perspective, our work extends the
aforementioned derivation to a broader class of circuits.

\subsection{Related ZNE advances}
\label{sec:level2.5}

Reference~\cite{Mohammadipour_2025} analyzes polynomial ZNE, deriving tight bias and variance bounds and showing that simple polynomial or exponential models can incur large approximation errors when the noise channel deviates from pure exponential behavior.

Reference~\cite{miranskyy2026improvingzeronoiseextrapolationphysically} introduces physically bounded extrapolation that constrains the zero-noise estimate to the physical range of observables, which can be applied as a post-processing step on our hybrid model.

Reference~\cite{pal2026foldingfreezeronoiseextrapolationlayoutinduced} proposes folding-free ZNE (FF-ZNE), which uses layout-induced noise diversity across isomorphic hardware configurations to eliminate circuit folding. FF-ZNE requires access to multiple distinct hardware mappings, while our method works on a single device using circuit folding and targets periodic circuits whose noise distribution is approximately log-normal.

These three approaches address different problems (theoretical bounds, physical constraints, folding elimination). We focus on the theoretical foundation of our model and benchmark it against the ZNE methods discussed in the earlier subsections.

\section{Hybrid Gaussian Exponential Extrapolation Model}
\label{sec:level3.0}
Pauli noise is widely used due to its diagonal representation
in the Pauli basis. Arbitrary noise channels can be rendered as
Pauli noise via the Pauli twirling technique \cite{PhysRevA.94.052325}.
Starting from the Pauli noise channel representation,
we derive an exact expression for the expectation values of arbitrary Pauli observables.
For circuits with a periodic structure, we propose a
hybrid Gaussian-exponential extrapolation model under certain assumptions,
which shows improved error mitigation performance over previous models.
\begin{align}
    \langle P_{\beta} \rangle(k) = a_{0} e^{c_{2}k^{2}+c_{1}k} + (a_{1} + b k) e^{c_{1}k}, \label{eq0}
\end{align}
where $k$ is the amplification factor.

This section develops the hybrid Gaussian-exponential model in three steps.
\begin{itemize}
    \item \textbf{Step 1(Sec.~\ref{sec:level3.1}):}
    We represent the output of a noise circuit as a sum over Pauli transfer paths $\alpha$, each with a coefficient $ F_{\alpha} $ (ideal transfer) and a noise factor $W_{\alpha}$ (attenuation). The expectation value becomes a weighted sum over paths.

    \item \textbf{Step 2(Sec.~\ref{sec:level3.2}):}
    By extracting an approximate Markov process for the transfer of
Pauli operators, we prove that $\ln W $ follows a normal distribution, as the number of periods grows(central limit theorem).

    \item \textbf{Step 3(Sec.~\ref{sec:level3.3}):}
    Using the log-normal property and expanding the covariance between $ F $ and $\ln{W} $ to first order, we obtain the fitting model in Eq.~\ref{eq0}.
\end{itemize}

\subsection{Exact model}
\label{sec:level3.1}
For a given quantum circuit, we can partition it into $m$ modules
according to its structural characteristics.
Let $C_{k}$ denote the $k$\text{-th} module, and let $A^{(k)}$
be its adjacency structure matrix, where the entry $A^{(k)}_{i,j}$
is defined as $1$ if $\operatorname{tr}\bigl( P_{j} C_{k}(P_{i}) \bigr) \neq 0$,
i.e., the Pauli operator $P_{i}$ has a non-zero component on $P_{j}$
after propagation through $C_{k}$. Otherwise, $A^{(k)}_{i,j}=0$,
if all $A^{(k)}$ are identical, i.e., $A^{(k)}=A$ for all $k$, we
define the circuit as structurally periodic. If the circuit
in each period is exactly the same, it is called a periodic circuit.

Let $\mathcal{E}^{i}$ denote the Pauli noise channel for the
$i$\text{-th} period. In the Pauli basis $\mathcal{P}= \{P_{i} \}_{i=0,\dots,4^{n}-1}$,
it is represented as
\begin{align}
\mathcal{E}^{i}(\rho) = (1 - p^{(i)})\rho + \sum_{j\ne 0}p_{j}^{(i)}P_{j}\rho P_{j}^{\dagger},
\end{align}
where $P_{0}$ is the identity operator $\operatorname{I}$.

For $\mathcal{E}^{i}$, the set of Pauli operators $\mathcal{P}$
forms a complete set of eigenvectors. The eigenvalue corresponding
to $P_{j}$ is denoted by $\lambda^{(i)}_{j}$, where
\begin{align}
\lambda_{j}^{(i)} &= \sum_{k:[P_{j},P_{k}]=0}p_{k}^{(i)} - \sum_{k:\{P_{j},P_{k}\}=0}p_{k}^{(i)} \nonumber\\
&= 1 - 2\sum_{k:\{P_{j},P_{k}\}=0}p_{k}^{(i)}.
\end{align}

In the Pauli basis, $C_{i}$ is expressed as
\begin{align}
C_{i}(\rho) = \sum_{j,k} C^{i}_{k,j} P_{j} \frac{\operatorname{tr}(\rho P_{k}^{\dagger})}{2^{n}},
\end{align}
where $C^{i}_{k,j} = \frac{\operatorname{tr}\bigl( P_{j} C_{i}(P_{k}) \bigr)}{2^{n}}$
is the projection coefficient of $P_{j}$ after $C_{i}$ acts on Pauli operator $P_{k}$.

For the $i$\text{-th} period, under Pauli twirling, the forward and backward noise channels are given respectively by
\begin{align}
    C_{i,\text{noise}} &= \mathcal{E}^{i}_{f} \circ C_{i},\\
    C^{\dagger}_{i,\text{noise}} &= C^{\dagger}_{i} \circ \mathcal{E}^{i}_{b} \label{eq6}.
\end{align}
In general, $\mathcal{E}^{i}_{f} \ne \mathcal{E}^{i}_{b}$. When the circuit is folded
$r$ times, the noise channel becomes
\begin{align}
    C^{r}_{i,\text{noise}} = (\mathcal{E}^{i}_{f} \circ \mathcal{E}^{i}_{b})^{r} \circ \mathcal{E}^{i}_{f} \circ C_{i} = \mathcal{E}^{i}_{r} \circ C_{i} \label{eq7}.
\end{align}
The eigenvalues of $C^{r}_{i,\text{noise}}$ are given by $\lambda_{j,r}^{(i)}$
for each Pauli operator $P_{j}$, with
\begin{align}
\lambda_{j,r}^{(i)} = (\lambda_{j,f}^{(i)} \lambda_{j,b}^{(i)})^{r}\lambda_{j,f}^{(i)}\label{eq8},
\end{align}
where $\lambda_{j,f}^{(i)}$ and $\lambda_{j,b}^{(i)}$
are the corresponding eigenvalues of $\mathcal{E}^{i}_{f}$
and $\mathcal{E}^{i}_{b}$, respectively.

Let the initial state be $\rho^{0} = \sum_{i}\rho_{i}P_{i}$.
After propagating through $C_{1}$, the state transforms as
\begin{align}
    \rho^{1} = C_{1}(\rho^{0}) &= \sum_{j,k} C^{1}_{k,j} P_{j} \frac{\operatorname{tr}\bigl( \sum_{i} \rho_{i} P_{i} P_{k}^{\dagger} \bigr)}{2^{n}} \nonumber\\
    &= \sum_{j} \biggl( \sum_{k} C^{1}_{k,j} \rho_{k} \biggr) P_{j} \label{eq9}.
\end{align}
Applying the noise channel $\mathcal{E}^{1}_{r}$ yields
\begin{align}
    \mathcal{E}^{1}_{r}(\rho^{1}) = \sum_{j} \lambda_{j,r}^{(1)} \biggl( \sum_{k} C^{1}_{k,j} \rho_{k} \biggr) P_{j}.
\end{align}
After $m$ periods, the final output of the circuit is given by
\begin{align}
    C_{\text{noise}}(\rho^{0}) = \sum_{\alpha_{m}} \lambda_{\alpha_{m},r}^{(m)} \left( \sum_{\alpha_{m-1}} C^{m}_{\alpha_{m-1},\alpha_{m}} \lambda_{\alpha_{m-1},r}^{(m-1)} \dots \sum_{\alpha_{1}} C^{2}_{\alpha_{1},\alpha_{2}} \lambda_{\alpha_{1},r}^{(1)} \sum_{\alpha_{0}} C^{1}_{\alpha_{0},\alpha_{1}} \rho_{\alpha_{0}} \right) P_{\alpha_{m}} \label{eq11}.
\end{align}

Let $\alpha=(\alpha_{0}, \alpha_{1},\dots,\alpha_{m})$ be
a Pauli operator transfer path, with $F_{\alpha}$ the
associated transfer coefficient and $W_{\alpha}$ the noise coefficient, i.e., the attenuation factor along different Pauli transfer paths:
\begin{align}
    F_{\alpha} &= C^{m}_{\alpha_{m-1},\alpha_{m}} C^{m-1}_{\alpha_{m-2},\alpha_{m-1}} \cdots C^{1}_{\alpha_{0},\alpha_{1}}\rho_{\alpha_{0}} \label{eq12},\\
    W_{\alpha} &= \lambda_{\alpha_{m},r}^{(m)} \lambda_{\alpha_{m-1},r}^{(m-1)} \cdots \lambda_{\alpha_{1},r}^{(1)} \label{eq14}.
\end{align}
The expectation value of the Pauli observable $P_{\beta}$ for the
output of the ideal circuit is
\begin{align}
    \langle P_{\beta} \rangle_{0} = \operatorname{tr}\bigl( C_{\text{ideal}}(\rho^{0}) P_{\beta} \bigr) = 2^{n}\sum_{\alpha:\alpha_{m}=\beta} F_{\alpha} \label{eq13},
\end{align}
whereas for the noisy circuit
\begin{align}
    \langle P_{\beta} \rangle(r) = \operatorname{tr}\bigl( C_{\text{noise}}(\rho^{0}) P_{\beta} \bigr)
    = 2^{n}\sum_{\alpha:\alpha_{m}=\beta} F_{\alpha} W_{\alpha} \label{eq15}.
\end{align}

Define $D_{\beta} = \{ \alpha : \alpha_{m} = \beta \}$ as
the set of paths terminating at $P_{\beta}$, with $N_{\beta}=\#D_{\beta}$
the total number of paths. Regard $F$ and $W$
as random variables on $D_{\beta}$. Then Eqs.~\eqref{eq13}
and \eqref{eq15} can be viewed as equally weighted sums over a discrete index.
Let $\operatorname{unif}(D_{\beta})$ be the
uniform distribution over $D_{\beta}$. Consequently,
\begin{align}
    \langle P_{\beta} \rangle(r) &= 2^{n}N_{\beta} \, \mathbb{E}_{\alpha \sim \operatorname{unif}(D_{\beta})}[F W] \nonumber\\
    &= \langle P_{\beta} \rangle_{0} \, \mathbb{E}_{\alpha \sim \operatorname{unif}(D_{\beta})}[W]  \nonumber\\
    &\quad + 2^{n}N_{\beta} \operatorname{cov}_{\alpha \sim \operatorname{unif}(D_{\beta})}(F,W) \label{eq16}.
\end{align}

\subsection{Noise statistical analysis}
\label{sec:level3.2}
In this section, we demonstrate that, as the number of periods
tends to infinity, the noise scaling factor $W$ is
weakly approaching a log-normal distribution.

For two sequences of random variables, $\{Y_i\}_{i\ge 1}$
and $\{Z_i\}_{i\ge 1}$, with distribution laws
$\{\mathcal{L}(Y_i)\}_{i\ge 1}$ and
$\{\mathcal{L}(Z_i)\}_{i\ge 1}$, respectively,
if for every bounded continuous function $f(\cdot)$,
$\mathbb{E}(f(Y_n)) \xrightarrow{n\to\infty} \mathbb{E}(f(Z_n)),$
then $\{\mathcal{L}(Y_i)\}_{i\ge 1}$ and
$\{\mathcal{L}(Z_i)\}_{i\ge 1}$ are said to
have weakly approaching distribution laws, denoted as
$\{\mathcal{L}(Y_i)\}_{i\ge 1} \overset{wa}{\longleftrightarrow} \{\mathcal{L}(Z_i)\}_{i\ge 1}$\cite{article}.

\begin{theorem}\label{th1}
For a periodic circuit, let $A$ be the adjacency matrix of a single period and
$m$ be the number of periods. For a Pauli observable $P_{\beta}$,
let $W$ denote the noise coefficient defined by Eq.~\eqref{eq14}.
Assuming that the noise in each period is Pauli noise and that
$A$ is primitive (i.e., irreducible and aperiodic), then
\begin{align}
    \mathcal{L}\!\left(\frac{\ln W - \mathbb{E}(\ln W)}{\sqrt{m}}\right) \overset{wa}{\longleftrightarrow} \mathcal{N}\!\left(0,\frac{\operatorname{var}(\ln W)}{m}\right) \label{eq17}.
\end{align}
\end{theorem}

Consider a sequence of random variables $\mathbf{X} = \{X_k : k \in \mathbb{Z}\}$
defined on a probability space $( \Omega, \mathcal{F}, P )$.
Let $\mathcal{F}_{m}^{n}$ be the $\sigma$\text{-algebra}
generated by $\{X_k : m \le k \le n\}$.
The strong mixing coefficient is then defined as
\begin{align}
    \alpha(\mathbf{X}, k) = \sup_{n} \sup_{\substack{A \in \mathcal{F}_{-\infty}^{n},\\B\in \mathcal{F}_{n+k}^{\infty}}} |P(A \cap B) - P(A)P(B)| \label{eq18}.
\end{align}

The proof of Theorem~\ref{th1} requires the following central limit theorem.

\begin{lemma}[Central Limit Theorem \cite{EKSTROM2014236}]
\label{lemma1}
Let $\{X_{n,k} : 1 \le k \le d_n\}$ be a triangular array
of random variables, $S_{n, d_{n}}=\sum_{i=1}^{d_{n}} X_{n, i}$
and $S_{n, b, t}=\sum_{i=t}^{t+b-1} X_{n, i}$. Let
$\alpha_{n}(\cdot)$ be the strong mixing coefficient corresponding to the $n$th row.
Then
\begin{align}
    \mathcal{L}\!\left(\frac{S_{n,d_n}-\mathbb{E}S_{n,d_n}}{\sqrt{d_n}}\right) \overset{wa}{\longleftrightarrow} \mathcal{N}\!\left(0, \operatorname{var}(\frac{S_{n,d_n}}{\sqrt{d_n}})\right) \label{eq19}
\end{align}
holds true if for some $\delta>0$, the following conditions \upshape{(B1)} and \upshape{(B2)}
are satisfied.
\begin{enumerate}
    \item[\upshape (B1)] $\displaystyle \mathbb{E}\bigl[|X_{n,k} - \mathbb{E}X_{n,k}|^{2+\delta}\bigr] < c$ for some $c>0$ and all $n,k$;
    \item[\upshape (B2)] $\displaystyle \sum_{k=0}^{\infty} (k+1)^2 \alpha_n^{\frac{\delta}{4+\delta}}(k) < c$ for some $c>0$ and all $n$.
\end{enumerate}
\end{lemma}

\begin{lemma}[Perron--Frobenius Theorem \cite{Horn_Johnson_1985}]
\label{lemma2}
If $A\in M_{n}$ is nonnegative and primitive, then there exists a
unique largest eigenvalue $\lambda_{1}$ with $\lambda_{1}>0$, and
all other eigenvalues satisfy $|\lambda| < \lambda_1$.
If $\ket{r}$ and $\bra{l}$ are, respectively, the right and left
Perron vectors of $A$, then
\begin{align}
        A =\lambda_{1}\ket{r}\bra{l}+R,\label{eq20}
\end{align}
where $\bra{l}R=\bra{0}$, $R\ket{r}=\ket{0}$ and $\rho(R)<\lambda_{1}$.
Consequently,
\begin{align}
    A^{k}=\lambda_{1}^{k} \ket{r}\bra{l} + R^{k}.
\end{align}
\end{lemma}

By standard results in matrix analysis \cite{Horn_Johnson_1985}, for a
given matrix norm $\|\cdot\|$ and any $\epsilon>0$, there exists
a constant $C$ such that
\begin{align}
    \|R^k\| \le C (\rho(R) + \epsilon)^k \label{eq21}.
\end{align}

We now present the proof of the Theorem~\ref{th1}.

Let $S_{0}$ be the space of nonzero Pauli components of the initial state.
Define $f(t, i)=\sum_{s \in S_{0}}(A^{t})_{s,i}$ as the number
of paths reaching $P_{i}$ after $t$ periods, and $g(t,i)=(A^{m-t})_{i,\beta}$
as the number of paths from $P_{i}$ to $P_{\beta}$ through $(m-t)$ periods.
The total number of paths is $N=f(m,\beta)$. Under the uniform path distribution,
the marginal distribution at the $t$\text{-th} period is
\begin{align}
    \mu_{t}(i)=p(\alpha_{t}=i)=\frac{f(t,i)\,g(t,i)}{N}.\label{eq22}
\end{align}
The joint distribution probability for steps $t-1$ and $t$ is given by
\begin{align}
    p(\alpha_{t-1}=i,\alpha_{t}=j)=\frac{f(t-1,i)\,g(t,j)\,A_{i,j}}{N}\label{eq23}.
\end{align}
Consequently, the transition probability matrix, denoted by $P_{t}$,
takes the form
\begin{align}
    (P_{t})_{i,j}= \frac{p(\alpha_{t-1}=i,\alpha_{t}=j)}{p(\alpha_{t-1}=i)}=\frac{g(t,j)\,A_{i,j}}{g(t-1,i)}\label{eq24}.
\end{align}
To demonstrate that this process is a Markov process,
consider the joint probability
\begin{align}
    &\quad p(\alpha_{t}=i_{t},\dots,\alpha_{0}=i_{0}) \nonumber \\ &=\frac{f(0,i_{0})\,A_{i_{0},i_{1}}\cdots A_{i_{t-1},i_{t}}\,g(t,i_{t})}{N}.\label{eq25}
\end{align}
It follows that
\begin{align}
    &p(\alpha_{t}=i_{t}\mid \alpha_{t-1}=i_{t-1},\dots,\alpha_{0}=i_{0}) \nonumber\\
    &\qquad = \frac{p(\alpha_{t}=i_{t},\dots,\alpha_{0}=i_{0})}{p(\alpha_{t-1}=i_{t-1},\dots,\alpha_{0}=i_{0})} \nonumber\\
    &\qquad = \frac{g(t,i_{t})\,A_{i_{t-1},i_{t}}}{g(t-1,i_{t-1})} \nonumber\\
    &\qquad = p(\alpha_{t}=i_{t}\mid \alpha_{t-1}=i_{t-1})\label{eq26}.
\end{align}

From Lemma~\ref{lemma2} and Eq.~\eqref{eq24}, we derive the
following expression for an approximately time-homogeneous Markov process:
\begin{align}
    (P_{t})_{i,j} &= A_{i,j}\,\frac{\lambda_{1}^{m-t}\langle j|r\rangle\langle l|\beta\rangle + \langle j|R^{m-t}|\beta\rangle}{\lambda_{1}^{m-(t-1)}\langle i|r\rangle\langle l|\beta\rangle + \langle i|R^{m-(t-1)}|\beta\rangle} \nonumber\\
    &= A_{i,j}\,\frac{r_{j}}{\lambda_{1}r_{i}} + O\!\left(\frac{\|R^{m-t}\|}{\lambda_{1}^{m-t}}\right)\label{eq27}.
\end{align}
Setting $P$ as the transition matrix defined by $P_{i,j}=A_{i,j}\,\frac{r_{j}}{\lambda_{1}r_{i}}$,
we obtain a time-homogeneous Markov process with stationary distribution
$\pi$ given by $\pi_{i}=l_{i}r_{i}$.

Consider a sufficiently small $\epsilon>0$ and let
$\rho=\frac{|\lambda_{2}+\epsilon|}{\lambda_{1}}<1$. Define
$\Delta_{t}=P_{t}-P$. In the $l_{\infty}$ matrix norm, the error satisfies
$\|\Delta_{t}\|_{\infty}=O(\rho^{m-t})$. The marginal distribution can
then be expressed as
\begin{align}
    \mu_{t}(i)=\pi_{i}+O(\rho^{t}+\rho^{m-t})\label{eq28}.
\end{align}
By Lemma~\ref{lemma2}, $P$ admits a similar decomposition
\begin{align}
    P=\ket{\mathbf{1}} \bra{\pi}+R_{1}\label{eq30},
\end{align}
with the same structure as $A$, where $\|R_{1}^{k}\|_{\infty}=O(\rho^{k})$.

Let $P_{t,k}$ denote the transition matrix for $k$ steps beginning
at step $t$ and set $R_{t,k}=P_{t,k}-P^{k}$. Then
\begin{align}
    \|R_{t,k}\|_{\infty} &= \|P_{t+k-1}\cdots P_{t+1} - P^{k}\|_{\infty} \nonumber\\
    &= \Bigl\|\sum_{s=0}^{k-1} P_{t+k-1}\cdots P_{t+s+1}\,\Delta_{t+s}\,P^{s}\Bigr\|_{\infty} \nonumber\\
    &\le \sum_{s=0}^{k-1} \|\Delta_{t+s}\|_{\infty} \nonumber\\
    &= O(\rho^{m-t-k}) \label{eq31}.
\end{align}

To simplify the analysis, we split the $m$ periods into two subsets. Let
$b_m = \lfloor c \ln m \rfloor$ with $c=-\frac{\alpha}{\ln{\rho}}$.
Define the boundary as $\partial_m \doteq \{i : 0 \le i < b_m \ \text{or} \ m-d_m < i \le m\}$,
and the interior as $I_{m} \doteq \{i:b_{m} \le i \le m-b_{m} \}$.

From Eq.~\eqref{eq14}, we introduce the random variables $X_{i}=\ln{\lambda^{(i)}}$
and define $S_{m}=\ln{W}$, which satisfies $S_{m}=\sum_{i=1}^{m}X_{i}$.
The variance of $S_{m}$ can be expressed as
\begin{align}
    \operatorname{var}(S_m) = \sum_{i=1}^m \operatorname{var}(X_i) + 2 \sum_{1 \le i < j \le m} \operatorname{cov}(X_i, X_j). \label{eq32}
\end{align}

For the first term, set $v_i = \operatorname{var}_\pi(X_i)$. By Eq.~\eqref{eq28}, we have
\begin{align}
    \operatorname{var}(X_i) = v_i + O(\rho^{i} + \rho^{m-i}).
\end{align}
In a periodic circuit, the noise strength remains constant across periods.
When the noise is depolarizing, we obtain $\operatorname{var}(S_m) = 0$, in which case
Theorem~\ref{th1} is trivially fulfilled and the normal distribution
collapses to a single point. For non-depolarizing noise, we have
$v_{k} =\gamma_{0}$ with $\gamma_{0}>0$. Hence,
\begin{align}
\sum_{i=1}^{m} \operatorname{var}(X_{i}) = m\gamma_{0} + O(1).
\end{align}

For the latter term, we consider the covariance of distinct components:
\begin{align}
    \operatorname{cov}(X_t, X_{t+k}) = \mathbb{E}(X_t X_{t+k}) - \mathbb{E}(X_t)\,\mathbb{E}(X_{t+k}).
\end{align}
Let $M_t = \operatorname{diag}(X_t)$, then
\begin{align}
    \mathbb{E}(X_t) = \sum_{i=0}^{4^n-1} \mu_t(i) X_t(i) = \langle \mu_t | M_t | \mathbf{1} \rangle.
\end{align}
Consequently,
\begin{align}
    &\quad \operatorname{cov}(X_t, X_{t+k})  \nonumber\\
    &= \langle \mu_t | M_t P_{t,k} M_{t+k} | \mathbf{1} \rangle - \langle \mu_t | M_t | \mathbf{1} \rangle \langle \mu_{t+k} | M_{t+k} | \mathbf{1} \rangle \nonumber\\
    &= \bigl( \langle \mu_t | M_t - \langle \mu_t | M_t | \mathbf{1} \rangle \langle \mu_t | \bigr) P_{t,k} M_{t+k} | \mathbf{1} \rangle \nonumber\\
    &= \bigl( \langle \mu_t | M_t - \langle \mu_t | M_t | \mathbf{1} \rangle \langle \mu_t | \bigr) (R_1^k + R_{t,k}) M_{t+k} | \mathbf{1} \rangle \nonumber\\
    &\le 2 \| M_t \|_\infty \| \mu_t \|_1 \| M_{t+k} \|_\infty \bigl( \| R_1^k \|_\infty + \| R_{t,k} \|_\infty \bigr) \nonumber\\
    &= O(\rho^{m-t-k} + \rho^{k}).
\end{align}
For periodic circuits, we have $M_{t}=M$ for all $t>0$. In conjunction with Eq.~\eqref{eq28}, we set
\begin{align}
    \gamma_k = \langle \pi | M R_1^k M | \mathbf{1} \rangle,
\end{align}
which satisfies $\gamma_{k}=O(\rho^{k})$. It then follows that
\begin{align}
    \operatorname{cov}(X_t, X_{t+k}) = \gamma_k + \eta_{t,k},
\end{align}
where
\begin{align}
        \eta_{t,k} &= O(\rho^{t} + \rho^{m-t})  \| R_1^k + R_{t,k}\|_\infty   + O(\| R_{t,k} \|_\infty) \nonumber\\
    &= O(\rho^{t+k} + \rho^{m-t-k}).
\end{align}

In $S_{I_{m}}$,
\begin{align}
        &\sum_{b_m \le i < j \le m-b_m} \operatorname{cov}(X_i, X_j) \nonumber\\
    &= \sum_{k=1}^{m-2b_m} \sum_{t=b_m}^{m-b_m-k} \operatorname{cov}(X_t, X_{t+k}) \nonumber\\
    &= m \sum_{k=1}^{m-2b_m} \gamma_k + O(\ln m) + O(m^{1-\alpha}) \nonumber\\
    &= O(m) \label{eq41}.
\end{align}
By Eq.~\eqref{eq32} and Eq.~\eqref{eq41}, we obtain
\begin{align}
    \operatorname{var}(S_{I_m}) = \biggl( \gamma_0 + 2\sum_{k=1}^{m-2b_m} \gamma_k \biggr) m + O(\ln m+ m^{1-\alpha}) \label{eq42}.
\end{align}
Since $\gamma_{k}=O(\rho^{k})$, the first-order term is absolutely convergent.
Accordingly, we define
\begin{align}
    c = \lim_{m \to \infty} \left( \gamma_0 + 2\sum_{k=1}^{m-2b_m} \gamma_k \right).
\end{align}

Given that the noise from each individual gate is small, we have
$|X_{m,i}| = O(\lambda) \ll 1$, which readily fulfills
condition (B1) of Lemma~\ref{lemma1}. For arbitrary
$A \in \mathcal{F}_{-\infty}^{t}$ and $B \in \mathcal{F}_{t+k}^{\infty}$,
by virtue of the Markov property,
\begin{align}
    &\quad P(A \cap B \mid \alpha_t = i, \alpha_{t+k} = j) \nonumber\\
    &= P(A \mid \alpha_t = i, \alpha_{t+k} = j, B) \; P(B \mid \alpha_t = i, \alpha_{t+k} = j) \nonumber\\
    &= P(A \mid \alpha_t = i) \; P(B \mid \alpha_{t+k} = j).
\end{align}
Consequently,
\begin{align}
    &\quad \bigl| P(A \cap B) - P(A)P(B) \bigr| \nonumber\\
    &= \Bigl| \sum_{i,j}\bigl[ \bigl( P(\alpha_t = i, \alpha_{t+k} = j) -
    P(\alpha_t = i)P(\alpha_{t+k} = j) \bigr)  \nonumber\\
    &\qquad \times P(A \mid \alpha_t = i)P(B \mid \alpha_{t+k} = j) \bigr] \Bigr| \nonumber\\
    &\le \sum_{i,j} \bigl| P(\alpha_{t+k} = j \mid \alpha_t = i) - P(\alpha_{t+k} = j) \bigr| \nonumber\\
    &= O(\rho^{k} + m^{-\alpha}).
\end{align}
It follows that
\begin{align}
    &\sum_{k=0}^{\infty} (k+1)^2 \alpha_m^{\frac{\delta}{4+\delta}}(k) \nonumber \\
    =&\sum_{k=0}^{m-2b_m} (k+1)^2 O\!\left( (\rho^{\frac{\delta}{4+\delta}})^k + m^{-\frac{\delta}{4+\delta}\alpha} \right).
\end{align}
When $\frac{\delta}{4+\delta}\alpha \ge 3$, the above series converges absolutely.
In fact, choosing $\delta=4$ and $\alpha=6$ fulfills this requirement. Thus,
condition (B2) of Lemma~\ref{lemma1} holds. By Lemma~\ref{lemma1}, we have
\begin{align}
    \mathcal{L}\!\left( \frac{S_{I_m} - \mathbb{E}(S_{I_m})}{\sqrt{m-2b_m}} \right) \overset{wa}{\longleftrightarrow} \mathcal{N}\!\left(0, \frac{\operatorname{var}(S_{I_m})}{m-2b_m}\right) \label{eq47}.
\end{align}

On the boundary $\partial_{m}$, we obtain $|S_{\partial_m}| \le |\partial_m| \max_{1\le t\le m,\,1\le i\le 4^{n}-1} \bigl( X_t(i) \bigr) = O(\ln m)$, and
\begin{align}
        \operatorname{var}(S_{\partial_m}) &= \sum_{i \in \partial_m} \operatorname{var}(X_i) + \sum_{\substack{i,j \in \partial_m \\ i \ne j}} \operatorname{cov}(X_i, X_j) \nonumber\\
    &= O(\ln m).
\end{align}
Consequently,
\begin{align}
        \operatorname{var}(S_m) &= \operatorname{var}(S_{\partial_m}) + \operatorname{var}(S_{I_m}) + 2\operatorname{cov}(S_{\partial_m}, S_{I_m}) \nonumber\\
    &= \operatorname{var}(S_{I_m}) + O(\ln m)  \nonumber \\
    &\quad+ 2\rho_{\partial_m, I_m} \sqrt{\operatorname{var}(S_{\partial_m}) \operatorname{var}(S_{I_m})} \nonumber\\
    &= \operatorname{var}(S_{I_m}) + O(\ln m + \sqrt{m \ln{m}}). \label{eq49}
\end{align}
When $c=0$, the Gaussian part converges to a point mass in the limit,
and the extension of $S_{I_{m}}$ in Eq.~\eqref{eq47} to $S_{m}$ does not affect
the convergence result. For $c>0$, the boundary part
is negligible, as expressed by
\begin{align}
    \frac{S_{\partial_m} - \mathbb{E}[S_{\partial_m}]}{\sqrt{\operatorname{var}(S_m)}} \overset{P}{\longrightarrow} 0. \label{eq50}
\end{align}
Combining Eqs.~\eqref{eq47}, \eqref{eq49}, and \eqref{eq50}, together with Lemma 7 in
Belyaev and Sj\"ostedt-de Luna (2000) \cite{article}, we obtain Theorem~\ref{th1}.

Note that periodicity of the circuit is not essential in the proof.
For structurally periodic circuits, the central limit theorem continues
to hold, though with no guarantee of convergence for
the covariance of the normal component.

Theorem~\ref{th1} assumes that $A$ is primitive (irreducible and aperiodic). Irreducibility means the Pauli transfer graph is strongly connected: any Pauli operator can reach any other through a sequence of circuit periods. When the observable is a single Pauli operator, the Pauli operators traversed along all possible paths lie in the same connected space, which satisfies the requirement.

Aperiodicity requires that at least one diagonal entry of some power of $A$ be nonzero, which holds for generic circuits. Taking the Ising model as an example, each period comprises $RX$ and $RZ$ gates, together with pairs of $CNOT$ gates. For generic random parameter $\theta$, the $RX(\theta)$ and $RZ(\theta)$ gates are almost surely non-Clifford and, simultaneously, induce self-loops, that is, the output retains a nonzero overlap with the input. The paired $CNOT$ gates subsequently leave this self-overlap invariant, so that self-loops persist in the overall evolution. Due to the randomness of the noise, these self-loops cannot coherently interfere to exactly zero. Consequently, the entire circuit is aperiodic.

When $A$ is not primitive (e.g., circuits with exact Pauli selection rules), the Markov chain decomposes into communicating classes, each satisfying the CLT separately with class-conditional parameters.

If the observable $O=\sum_{\beta \in \mathcal{B}}c_{i}P_{i}$
is a non-Pauli observable, then according to Eq.~\eqref{eq15} we have
\begin{align}
        \langle O \rangle(r) &= \sum_{\beta \in \mathcal{B}} c_\beta \sum_{\alpha:\alpha_m = \beta} F_\alpha W_\alpha \nonumber\\
    &= \sum_{\alpha:\alpha_m \in \mathcal{B}} (c_\beta F_\alpha) W_\alpha .
\end{align}
This affects the distribution of the endpoints, which in turn
alters the stationary distribution. It does
not affect the validity of our proof. The method
can therefore be extended to general observables.

\subsection{Approximate Model}
\label{sec:level3.3}
In Appendix A of Ref.\cite{Jin_2024},
the relationship between the forward noise and the backward noise is given by
\begin{align}
    \lambda_{b,i}=\lambda_{f,i}e^{\lambda^{2}w_{i}},
\end{align}
where $\lambda$ is the noise strength per period, and
$w_{i}$ is a coefficient characterizing the second-order asymmetry
between the forward and backward noise on $P_{i}$. From Eqs.\eqref{eq8} and
\eqref{eq14}, we obtain
\begin{align}
        \ln W(r) &= (2r+1) \sum_{i=1}^{m} \ln \lambda^{(i)}_{f} + r \lambda^{2} \sum_{i=1}^{m} w^{(i)} \nonumber\\
    &= (2r+1) S_{m,0} + r \lambda^{2} S_{m,1}.
\end{align}

Theorem~\ref{th1} implies that $\ln{W(r)}$ follows a Gaussian distribution
$\mathcal{N}({\mu(r),\sigma^{2}(r)})$. In addition, $W(r)$ obeys a
log-normal distribution. Setting $k=2r+1$, we obtain
\begin{align}
    \mu(k) &= k \mu_{0} + \frac{k-1}{2} \lambda^{2} \mu_{1}, \\
    \sigma^{2}(k) &= k^{2} \sigma^{2}_{0} + \left( \frac{k-1}{2} \right)^{2} \lambda^{4} \sigma^{2}_{1} + k(k-1) \rho_{0,1} \lambda^{2} \sigma_{0} \sigma_{1},
\end{align}
where $\mu_{i}$ and $\sigma_{i}$ represent the expectation and variance of
$S_{m,i}$, respectively, and $\rho_{0,1}$ denotes their correlation coefficient.
The asymptotic scalings are given by $\mu_{0}=O(m\lambda)$,
$\sigma_{0}^{2}=O(m\lambda^{2})$, $\mu_{1}=O(m)$ and $\sigma_{1}^{2}=O(m)$.
By the properties of the log-normal distribution,
\begin{align}
    \mathbb{E}[W(k)] &= e^{\mu(k) + \frac{\sigma^{2}(k)}{2}}, \\
    \operatorname{Var}[W(k)] &= \bigl( e^{\sigma^{2}(k)} - 1 \bigr) e^{2\mu(k) + \sigma^{2}(k)}.
\end{align}
Combining with Eq.~\eqref{eq16} gives
\begin{align}
    \langle P_{\beta} \rangle = \left( \langle P_{\beta} \rangle_{0} + 2^{n}N_{\beta} \rho_{F,W} \sigma_{F} \sqrt{e^{\sigma^{2}}-1} \right) e^{\mu + \frac{\sigma^{2}}{2}}.
\end{align}
If $\rho_{F,W}$ is independent of $k$ and $\lambda_{f}=\lambda_{b}$
holds under the standard amplification factor, then
\begin{align}
    \langle P_{\beta} \rangle(k) = \left( \langle P_{\beta} \rangle_{0} + b \sqrt{e^{k\sigma^{2}} - 1} \right) e^{k\mu + \frac{k\sigma^{2}}{2}},
\end{align}
with $\langle P_{\beta} \rangle_{0}$, $b$, $\mu$, $\sigma^{2}$ to be fitted,
and the extrapolation taken to $k=0$. For a more precise error analysis,
we consider the Taylor expansion of $W(r)$. Defining $Z(k)=\ln{W(k)-\mu(k)}$,
which obeys $\mathcal{N}(0, \sigma^{2}(k))$, we have
$W(k)=e^{\mu(k)}\sum_{i=0}^{\infty}\frac{1}{i!}Z^{i}(k)$.
Substituting this expression into $\operatorname{cov}(F, W)$ leads to
\begin{align}
    \operatorname{cov}(F, W) = \sum_{i=1}^{\infty} \frac{1}{i!} \operatorname{cov}(F, Z^{i}(k)).
\end{align}
By truncating the expansion to order $s$, we can control the remainder error
and obtain a more accurate model. To reduce the number of fitting parameters,
we retain only the first-order term and control the remainder
under the condition $\sigma^{2}(k)<1/e$.

For $\ln{i!}$, we derive the following estimates via the rectangle rule
and the trapezoidal rule for the area, respectively:
\begin{align}
    \sqrt{2\pi} i^{i+\frac{1}{2}}e^{-i} \le i! \le ei^{i+\frac{1}{2}}e^{-i}.
\end{align}
Thus,
\begin{align}
    &\quad\sum_{i=2}^{\infty} \frac{1}{i!} \operatorname{cov}(F, Z^{i}(k))  \nonumber \\
    &\le \rho_{\max} \sigma_F \sum_{i=2}^{\infty} \frac{1}{i!} \sqrt{ \sigma^{2i}(k) \frac{(2i)!}{2^{i} i!} } \nonumber\\
    &\le \rho_{\max} \sigma_F \sum_{i=2}^{\infty} e^{\frac{1}{2}} \pi ^{-\frac{3}{4}}2^{-\frac{1}{2}} i^{-\frac{1}{2}} \left( \frac{2e \sigma^{2}(k)}{i} \right)^{\frac{i}{2}} \nonumber\\
    &\le \rho_{\max} \sigma_F \pi^{-\frac{1}{2}} 2^{-\frac{1}{2}} \sum_{i=2}^{\infty} \bigl( \sqrt{e \sigma^{2}(k)} \bigr)^{i} \nonumber\\
    &\le \rho_{\max} \sigma_F \pi^{-\frac{1}{2}} 2^{-\frac{1}{2}} \frac{e \sigma^{2}(k)}{1 - \sqrt{e \sigma^{2}(k)}}.
\end{align}
Combining with Eq.~\eqref{eq16} gives
\begin{align}
    \langle P_{\beta} \rangle = \langle P_{\beta} \rangle_{0} \, e^{\mu + \frac{\sigma^{2}}{2}} + 2^{n}N_{\beta} e^{\mu} \operatorname{cov}(F, Z) + \epsilon,
\end{align}
with $\epsilon(k) / \langle P_{\beta} \rangle(k) = O(\sigma^{2}) = O(m k^{2} \lambda^{2})$.
Let $\widetilde{P}_{\beta}(k) = \langle P_{\beta} \rangle(k) - \epsilon(k)$.
The following derivation shows that the relative error between the fitted
$\widetilde{P_{\beta}}(k)$ and $\langle P_{\beta} \rangle(k)$
remains of order $O(mk^{2}\lambda^{2})$.

Consider the true function expressed as
\begin{align}
    g(x)=f(x;a_{1}^{*},\dots,a_{n}^{*})+\epsilon(x).
\end{align}
Fitting $f$ with $n$ data points $(x_{i},g_{i})$ yields parameters
$\hat{a}=(\hat{a}_{1},\dots,\hat{a}_{n})$ satisfying $f(x_{i};\hat{a})=g(x_{i})$.
Setting $\delta a =\hat{a}-a^{*}$, we have the Taylor expansion
\begin{align}
    f(x_{i};\hat{a}) = f(x_{i};a^{*})+\sum_{j=1}^{n}\frac{\partial f(x_{i})}{\partial a_{j}}\delta a_{j} +O(\|\delta a\|^2),
\end{align}
from which it follows that
\begin{align}
    \sum_{j=1}^{n} \frac{\partial f(x_i)}{\partial a_j} \delta a_j + O(\|\delta a\|^2) = \epsilon(x_i).
\end{align}
Let $J_{i,j}=\frac{\partial f(x_{i})}{\partial a_{j}}$ and
$R(\delta a) = O(\|\delta a\|^2)$. Then
\begin{align}
    J\delta a + R(\delta a)=\epsilon.
\end{align}
This implies the inequality
\begin{align}
    \|J \delta a\| - \|R(\delta a)\| \le \|\epsilon(x)\| \le \|J \delta a\| + \|R(\delta a)\|,
\end{align}
and consequently $\|\delta a\| = O(\|\epsilon\|)$.
The resulting error under the fitted parameters $\hat{a}$ is
\begin{align}
    \Delta g &= f(x; \hat{a}) - \bigl( f(x; a^*) + \epsilon(x) \bigr) \nonumber\\
    &= \nabla_a f(x) \cdot \delta a - \epsilon(x).
\end{align}
Hence, $\Delta g/g = O(\epsilon/g)$.

Now consider the approximate model.
\begin{align}
\widetilde{P}_{\beta}
&= \langle P_{\beta} \rangle_{0} \, \exp\!\biggl( \mu + \frac{\sigma^{2}}{2} \biggr)
   + {2^{n}}N_{\beta} e^{\mu} \operatorname{cov}(F, Z) \nonumber\\
&= \langle P_{\beta} \rangle_{0} \, \exp\!\biggl(
   k^{2}\Bigl( \frac{\sigma_{0}^{2}}{2} + \frac{\lambda^{4}\sigma_{1}^{2}}{8} + \frac{\rho_{0,1}\lambda^{2}\sigma_{0}\sigma_{1}}{2} \Bigr) \nonumber \\
&\qquad + k\Bigl( \mu_{0} + \frac{\lambda^{2}\mu_{1}}{2} - \frac{\rho_{0,1}\lambda^{2}\sigma_{0}\sigma_{1}}{2} - \frac{\lambda^{4}\sigma_{1}^{2}}{4} \Bigr) \nonumber \\
&\qquad + \Bigl( -\frac{\lambda^{2}\mu_{1}}{2} + \frac{\lambda^{4}\sigma_{1}^{2}}{8} \Bigr) \biggr) \nonumber \\
&\quad + {2^{n}}N_{\beta} \Bigl( k \operatorname{cov}(F, S_{m,0}) + \frac{k-1}{2} \lambda^{2} \operatorname{cov}(F, S_{m,1}) \Bigr)  \nonumber\\
&\qquad \times \exp\!\biggl( k \Bigl(\mu_{0} + \frac{\lambda^{2}}{2}\mu_{1}\Bigr) - \frac{\lambda^{2}}{2}\mu_{1} \biggr).
\end{align}
Define
\begin{align}
\left\{\begin{matrix}
\begin{aligned}
 &c_{2}=\frac{\sigma_{0}^{2}}{2}+\frac{\lambda^{4}\sigma_{1}^{2}}{8} +\frac{\rho_{0,1}\lambda^{2}\sigma_{0} \sigma_{1}}{2},\\
 &c_{1}=\mu_{0} +\frac{\lambda^{2}\mu_{1}}{2} -\frac{\rho_{0,1} \lambda^{2}\sigma_{0} \sigma_{1}}{2} -\frac{\lambda^{4}\sigma_{1}^{2}}{4},\\
 &c_{0}=-\frac{\lambda^{2}\mu_{1}}{2}+\frac{\lambda^{4}\sigma_{1}^{2}}{8},\\
 &{c_{1}}'=\mu_{0}+\frac{\lambda^{2}}{2}\mu_{1},\\
 &{c_{0}}'=-\frac{\lambda^{2}}{2}\mu_{1}.
\end{aligned}
\end{matrix}\right.
\end{align}
Here $c_{2}=O(m\lambda^{2})$, $c_{1}=O(m\lambda)$, $\Delta c_{1}=c_{1}-{c_{1}}'=O(m\lambda^{3})$, $c_{0}=O(m\lambda^{2})$, $\Delta c_{0}=c_{0}-{c_{0}}'=O(m\lambda^{4})$.
When $\lambda\ll1$, we have $\Delta c_{i}/c_{i} =O(\lambda^{2})\ll1$ for $i=1,2$.
Thus, $c_{i}$ and ${c_{i}}'$ can be merged into a single coefficient. Besides,
$c_{0}$, as a multiplicative factor in the constant term, can be
absorbed into the other constant factors. Consequently, we set
\begin{align}
    \left\{\begin{matrix}
    \begin{aligned}
        &a_{0} = \langle P_{\beta} \rangle_{0} e^{c_{0}},\\
        &a_{1} = -\frac{{2^{n}}N_{\beta} \lambda^{2} \operatorname{cov}(F, S_{m,1})}{2} e^{c_{0}'},\\
        &b = {2^{n}}N_{\beta} \left( \operatorname{cov}(F, S_{m,0}) + \frac{ \lambda^{2} \operatorname{cov}(F, S_{m,1})}{2} \right) e^{c_{0}'}.
    \end{aligned}
    \end{matrix}\right.
\end{align}
We thus obtain the model
\begin{align}
    \langle P_{\beta} \rangle(k) = a_{0} e^{c_{2}k^{2} + c_{1}k} + (a_{1} + b k) e^{c_{1}k}. \label{eq93}
\end{align}
By extrapolating to $k=0$, we obtain the approximate value
$\langle P_{\beta} \rangle_{\text{ext}} = 2^nN_{\beta} \mathbb{E}[F e^{-\frac{\lambda^{2} w_{i}}{2}}]$,
effectively suppressing the noise strength from $\lambda$ to
$O(\lambda^{2})$. To further reduce this error, $w_{i}$ can be
treated as a constant factor. One strategy is to extract this
factor from the exact theoretical predictions of
a few circuits that are structurally identical and share
the same observable. The resulting factor can then be applied
to the extrapolated values of all analogous circuits.
Alternatively, following the purity-assisted method,
we may treat $\langle P_{\beta} \rangle_{\text{ext}}$ as a new noisy value
and employ
\begin{align}
    1 &= p_{0} = \frac{1}{2^n}\sum_{\beta} \langle P_{\beta} \rangle_{0}^{2},\\
    p_{\text{noise}} & \approx \frac{1}{2^n}\langle P_{0} \rangle_{0}^{2} + e^{-\lambda^{2} \bar{w}_{i}} \frac{1}{2^n}\sum_{\beta \neq 0} \langle P_{\beta} \rangle_{0}^{2}.
\end{align}
By measuring the purity, we may obtain the average attenuation coefficient.
However, this may introduce additional complications. The precise
measurement of purity is exponentially complex.
Although shadow tomography can yield unbiased estimates with
polynomial complexity, it still incurs substantial circuit overhead.
We therefore use the first method in our experiment.

In all experiments, the parameters in Eq.~\ref{eq93} are
obtained by least-squares fitting using the Levenberg-Marquardt
algorithm(as implemented in scipy.optimize.curve\_fit).
Since $c_{1}$ is dominated by the mean $\mu_0$, we set
$c_1 \le 0$. Since $c_2$ is strictly non-negative, we set
$c_2 \ge 0$, and no bounds are imposed on the other parameters.
The initial values are all chosen within $\left[-1, 1\right]$ and satisfy the boundary conditions of the parameters.

\section{Performance Comparison}
\label{sec:level4.0}
In this section, we compare the performance of standard ZNE,
IC-ZNE, pZNE, and Hybrid G-E ZNE on three classes of
circuits: the Trotterized dynamical evolution circuit
of a one-dimensional Ising spin chain, random circuits
and Grover circuits. To ensure reproducibility, we set
several random seeds using numpy.random.seed to generate
and run
circuits.

\subsection{One-dimensional transverse-field Ising model}
\label{sec:level4.1}
\begin{figure}[t]
\centering
\includegraphics[width=0.6\textwidth]{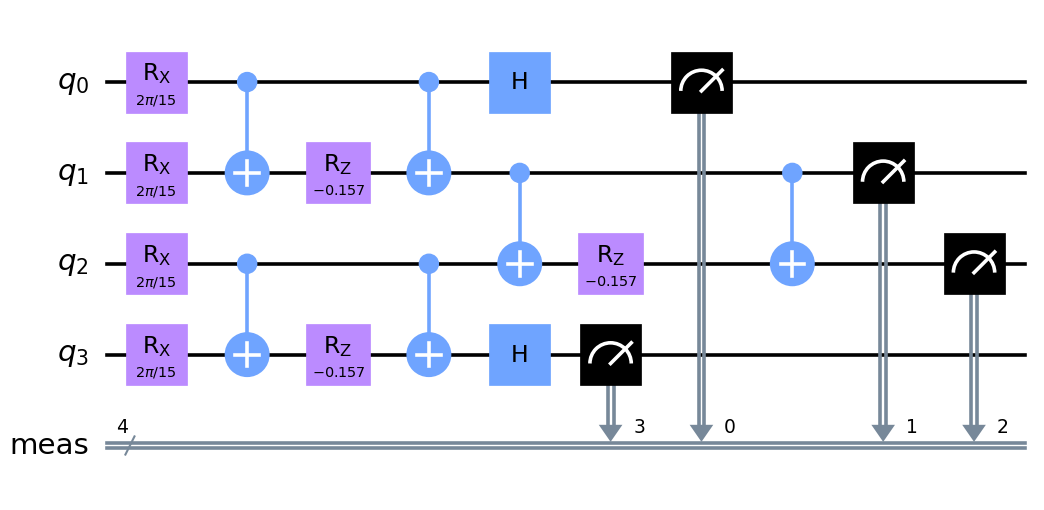}
\caption{\label{fig:step1} Schematic diagram of the circuit
for $J=0.37454$ and $dt = \pi /15$ with one Trotter step. Open boundary conditions are used for the 4-qubit chain.}
\end{figure}
For structured circuits, we first investigate the first-order
Trotterized dynamics of the 1D transverse-field Ising model with open boundary conditions.
The evolution is governed by the Hamiltonian
\begin{align}
    \hat{H}=-J\sum_{i=1}^{n-1}\hat{Z}_{i}\hat{Z}_{i+1}+h\sum_{i=1}^{n}\hat{X}_{i}\label{eq3.1},
\end{align}
where $J$ denotes the coupling strength between adjacent spins and
\begin{figure}[t]
    \centering
    \captionsetup{justification=raggedright, singlelinecheck=false}

    % (a)
    \begin{minipage}{0.48\linewidth}
        \centering
        \refstepcounter{subfig}\label{fig:a}
        \begin{tikzpicture}
            \node[inner sep=0] (img) {\includegraphics[width=\linewidth]{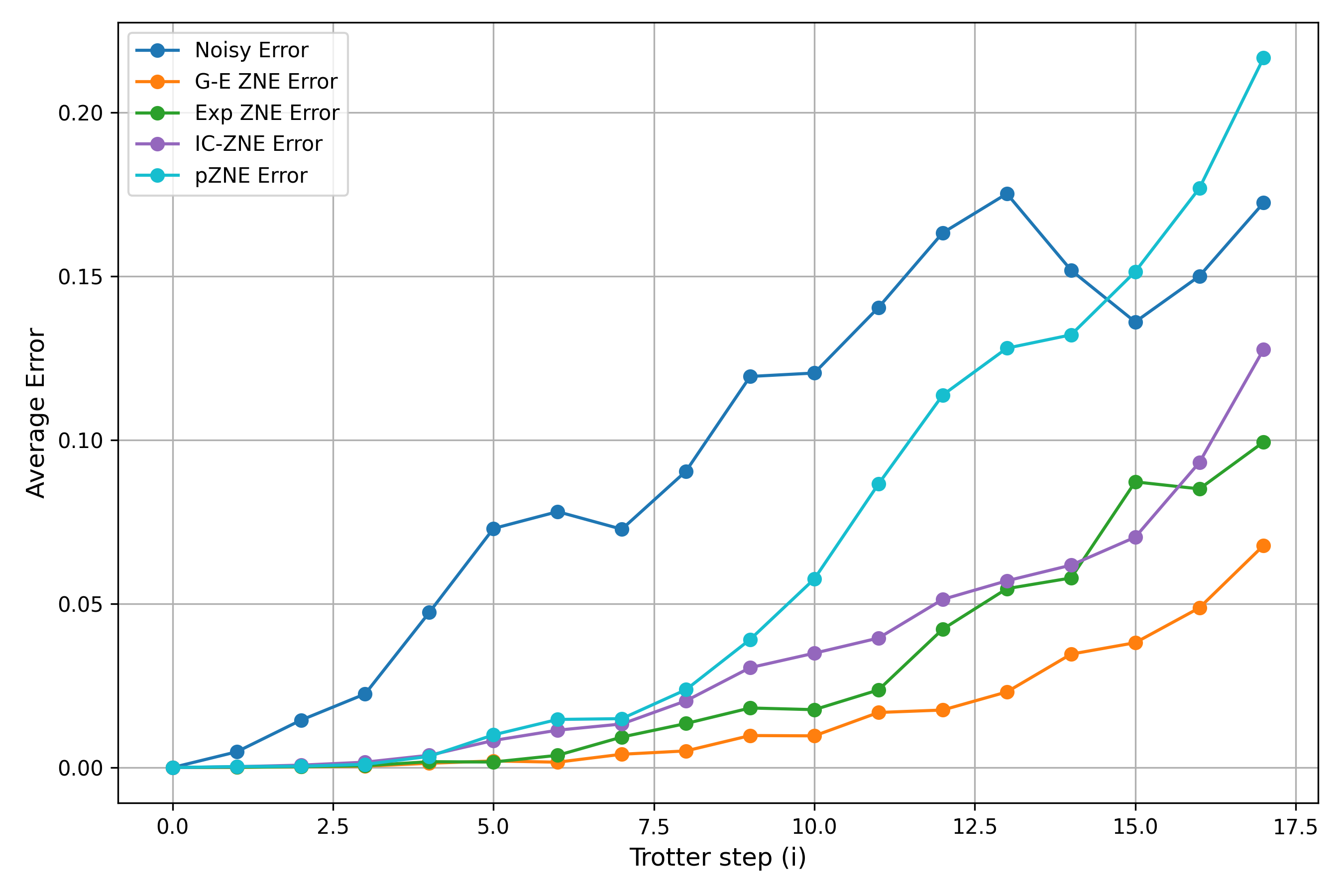}};
            \node[anchor=north west, font=\bfseries, inner sep=2pt, xshift=-6pt, yshift=-2pt] at (img.north west) {(a)};
        \end{tikzpicture}
    \end{minipage}
    \hspace{1em}
    % (b)
    \begin{minipage}{0.48\linewidth}
        \centering
        \refstepcounter{subfig}\label{fig:b}
        \begin{tikzpicture}
            \node[inner sep=0] (img) {\includegraphics[width=\linewidth]{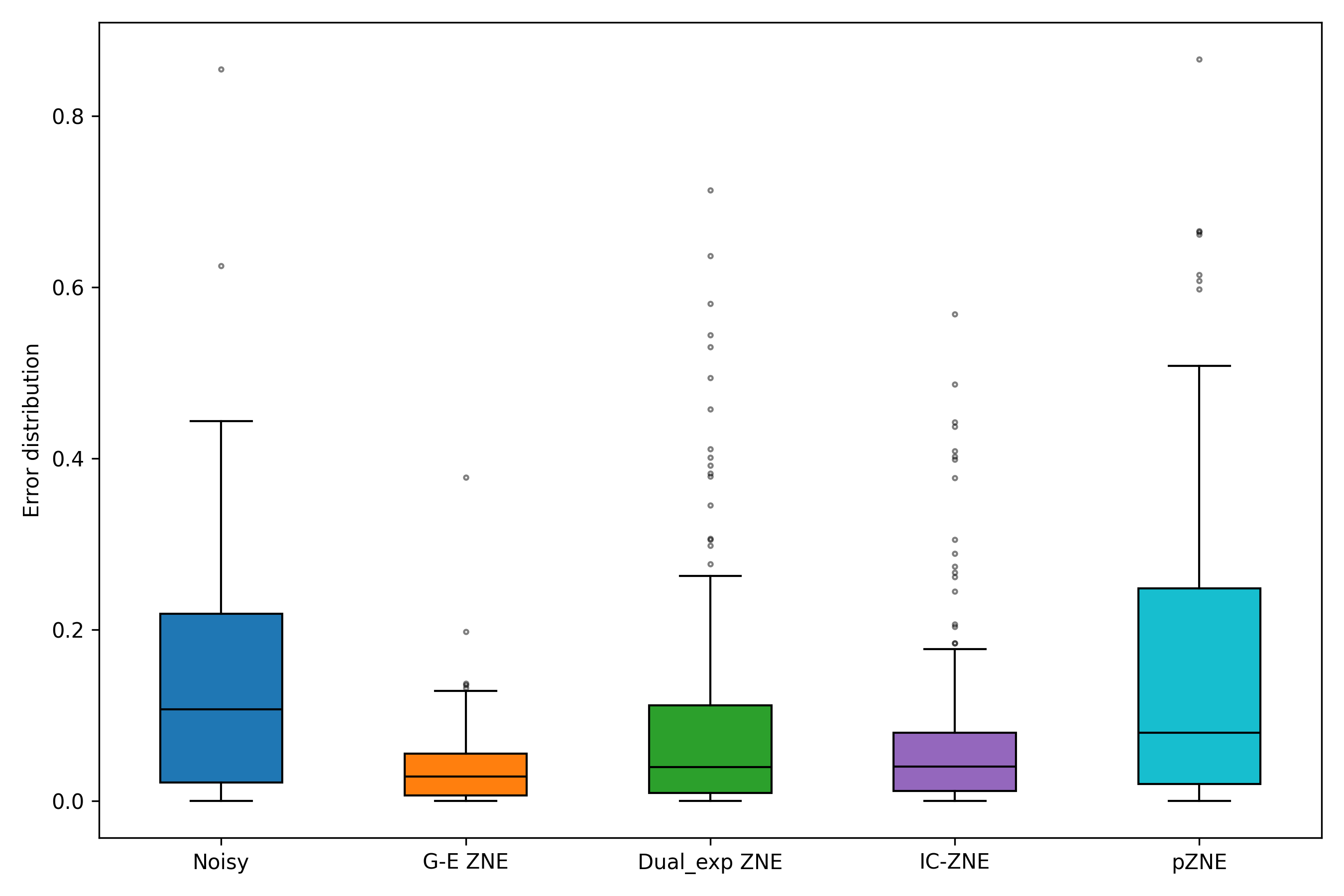}};
            \node[anchor=north west, font=\bfseries, inner sep=2pt, xshift=-6pt, yshift=-2pt] at (img.north west) {(b)};
        \end{tikzpicture}
    \end{minipage}

    \caption{Accuracy of different ZNE methods on the 1D transverse-field Ising model.
    (a). Mean absolute error versus circuit depth (number of steps)
    for different ZNE methods.
    (b). Box plots of absolute errors for different ZNE methods
    at depth 15 (200 circuits).
    The colored box spans the interquartile range (IQR), i.e.,
    the region between the first and third quartiles. The median is shown
    as a black horizontal line. Whiskers extend to 1.5 $\times$ IQR, and black
    dots denote outliers.}
    \label{fig:ising}
\end{figure}

$h$ the transverse magnetic field.
Fixing $h=1$, we uniformly
sample the coupling strength in the paramagnetic phase($J<h$)
to obtain 200 distinct Hamiltonians. Setting the time step per
Trotter layer to $dt=\pi/15$, we generate circuits with Trotter steps
ranging from $0$ to $17$ in unit increments.  Observables are
chosen uniformly from the set of 4-qubit Pauli observables.
Fig.\ref{fig:step1} depicts the circuit for $J=0.37454$ with
one Trotter step and the observable $P_{x}P_{z}P_{z}P_{x}$.

The experiments are carried out on IBM's FakeQuito simulator.
We employ the Pauli twirling technique to suppress non-Pauli noise
such as coherent noise. Noting that readout error is an
independent module with respect to ZNE, we obtain
expectation values directly from the density matrix
rather than via sampling. This eliminates the confounding
effect of readout error and gives a more precise characterization
of the model's inherent performance.

Fig.\ref{fig:ising}(a) shows the mean error of various methods across 200 circuits.
For fewer than 4 steps, where the
overall circuit noise is relatively low, all error
mitigation methods perform well. When the
number of steps exceeds 6, our results show
clear advantages, while the other methods gradually fail.
This indicates that our approach is applicable to a broader
range of noise levels and is particularly advantageous for
deeper circuits. Fig.\ref{fig:ising}(b) shows the distribution
of absolute errors obtained by different ZNE methods
across 200 circuits at $step=15$.
Our model shows better accuracy and stability than
the alternative extrapolation approaches.

\subsection{Random circuits}
\label{sec:level4.2}
In the second experiment, we benchmarked the performance of
random circuits. We generated 4-qubit random circuits
with two-qubit gate depths up to 36 in steps of 2,
and for each depth we generated 200 random circuits.

\begin{figure}[t]
\centering
\includegraphics[width=0.5\textwidth]{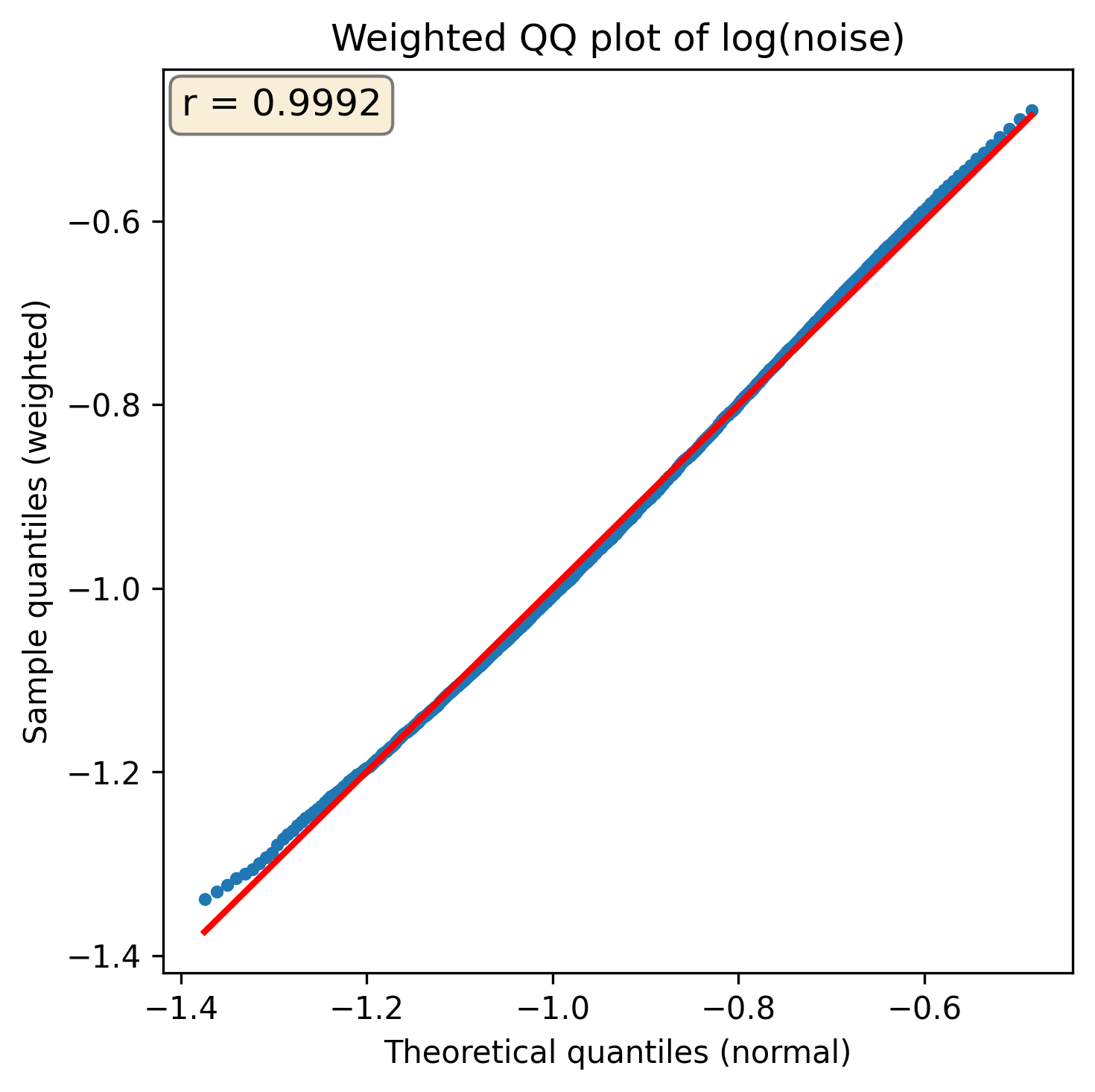}
\caption{\label{fig:qq_plot} Quantile-Quantile (Q-Q) plot at two-qubit depth of 18.}
\end{figure}

In the process of generating random circuits at a single
step, a comparable structure arises in expectation.
We sampled the noise scaling factors $(W_{\alpha})_{\alpha}$
over paths, obtaining $2$\text{e6} samples. As uniform sampling
over the path space is not feasible, we relied on weighted
sample quantiles. Using quantiles from 0.01 to 0.99 in steps
of 0.002, we generated a Q--Q plot of $\ln{W}$.
In Fig.\ref{fig:qq_plot}, the weighted quantile correlation
coefficient is 0.9992, indicating that $\ln{W}$ closely
follows a normal distribution.

\begin{figure}[t]
    \centering
    \captionsetup{justification=raggedright, singlelinecheck=false}

    % (a)
    \begin{minipage}{0.48\linewidth}
        \centering
        \refstepcounter{subfig}\label{fig:a2}
        \begin{tikzpicture}
            \node[inner sep=0] (img) {\includegraphics[width=\linewidth]{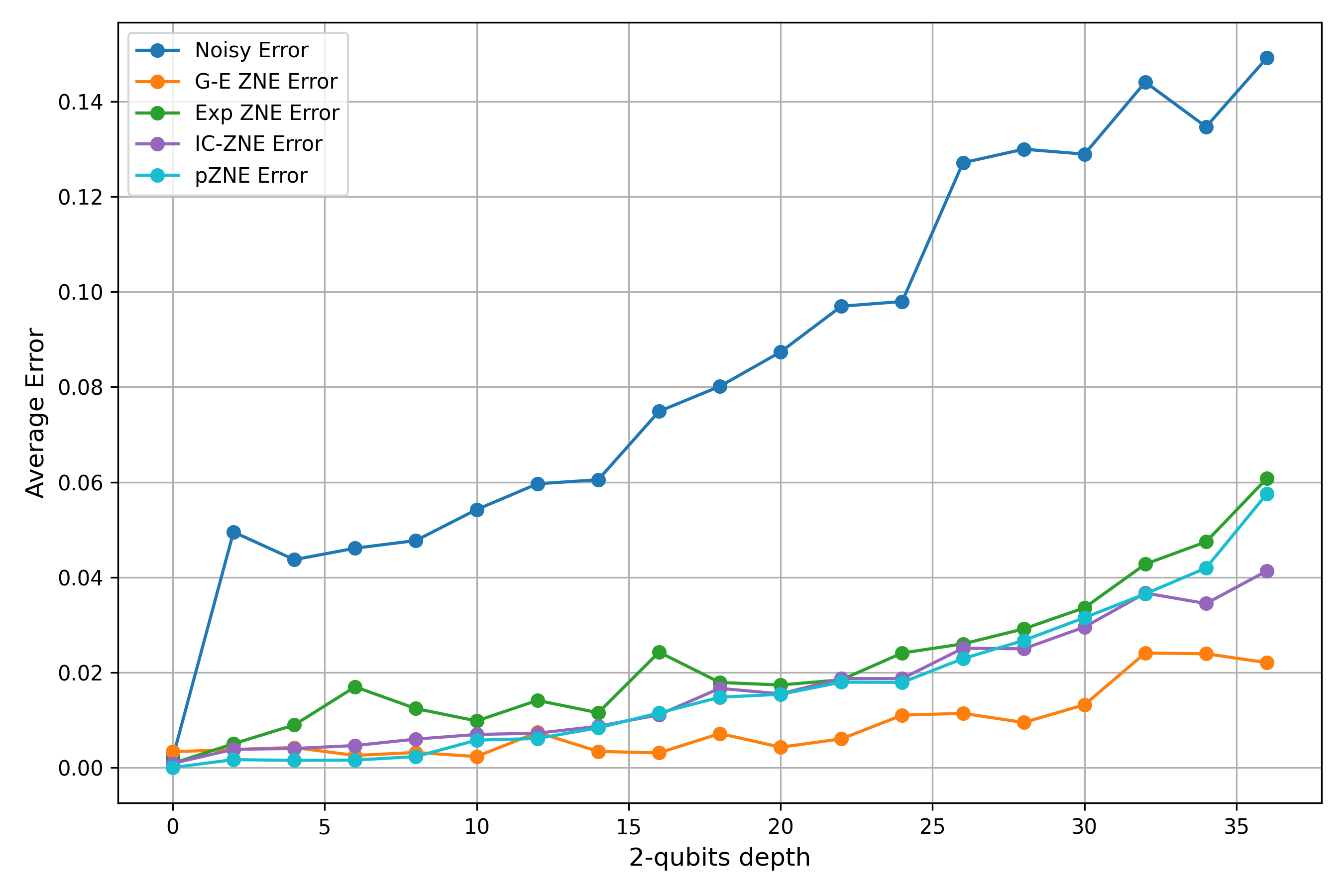}};
            \node[anchor=north west, font=\bfseries, inner sep=2pt, xshift=-6pt, yshift=-2pt] at (img.north west) {(a)};
        \end{tikzpicture}
    \end{minipage}
    \hspace{1em}
    % (b)
    \begin{minipage}{0.48\linewidth}
        \centering
        \refstepcounter{subfig}\label{fig:b2}
        \begin{tikzpicture}
            \node[inner sep=0] (img) {\includegraphics[width=\linewidth]{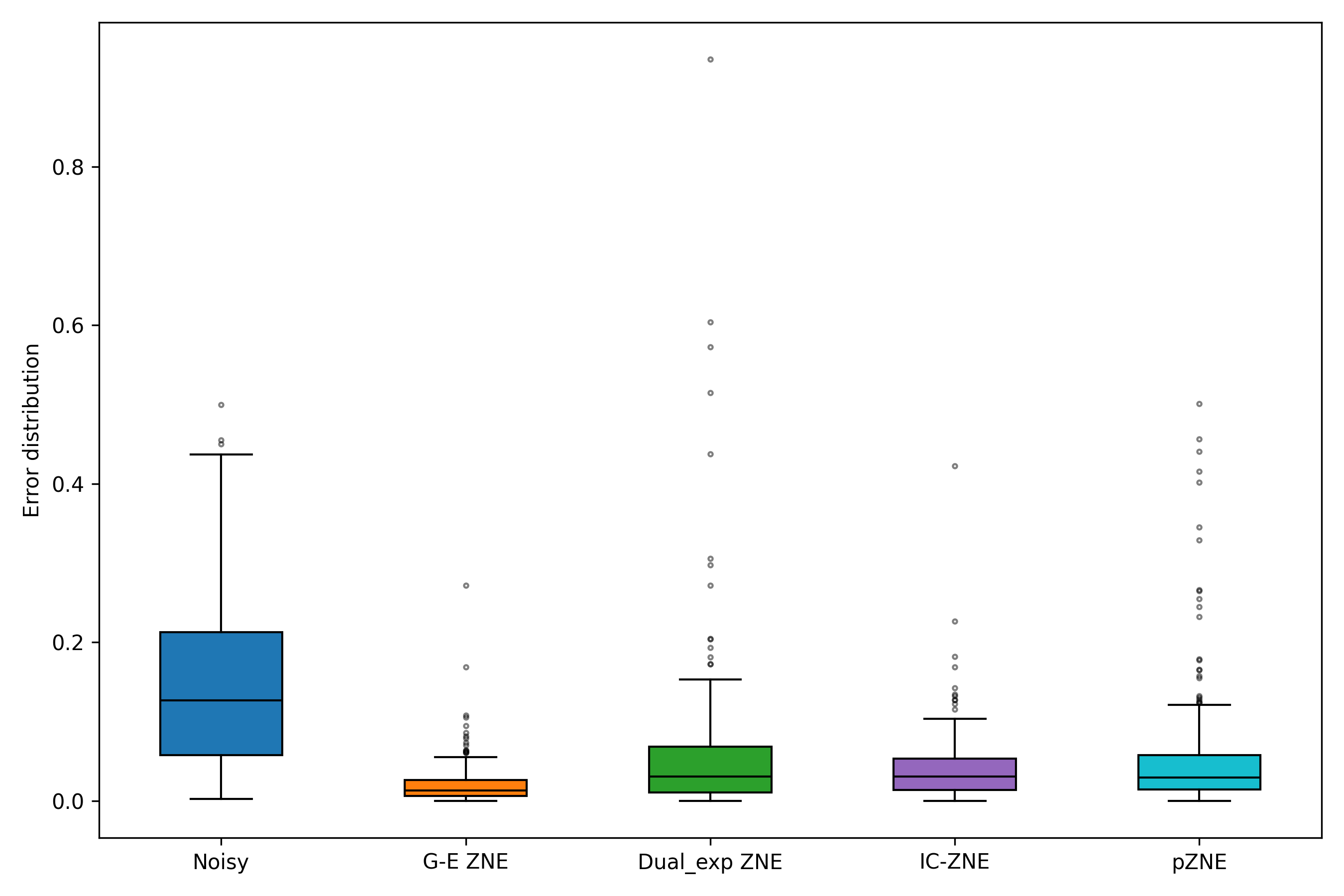}};
            \node[anchor=north west, font=\bfseries, inner sep=2pt, xshift=-6pt, yshift=-2pt] at (img.north west) {(b)};
        \end{tikzpicture}
    \end{minipage}

    \caption{Accuracy of different ZNE methods in random circuits.
    (a). Mean absolute error versus circuit depth (2-qubit gate depths)
    for different ZNE methods.
    (b). Box plots of absolute errors for different ZNE methods
    at depth 36 (200 circuits).}
    \label{fig:random}
\end{figure}

Fig.\ref{fig:random}(a) presents a comparison of different
error mitigation methods. Overall, for two-qubit gate depths
less than 8, our method is surpassed by pZNE. For depths
exceeding 12, it exhibits a clear advantage over the other
approaches within the scope of the experiment. This improvement
arises because, as the number of periods grows, the noise
increasingly follows a log-normal distribution,
which matches the basis of our model.
Fig.\ref{fig:random}(b)
shows the distribution
of absolute errors evaluated at $step=36$.
Our model shows better accuracy and stability than
the alternative extrapolation approaches.

\subsection{Grover circuits}
\label{sec:level4.3}
\begin{figure}[t]
\centering
\includegraphics[width=0.6\textwidth]{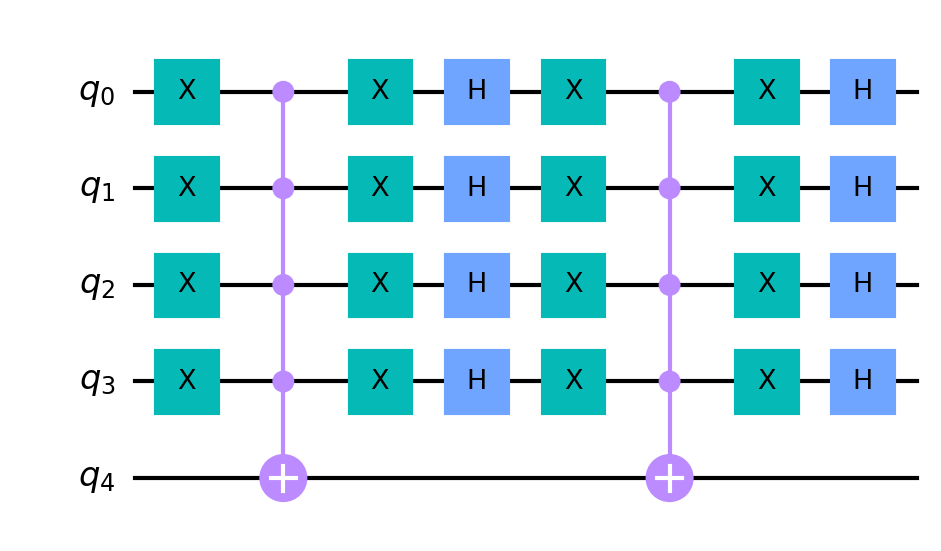}
\caption{\label{fig:grover_operator} Grover operator.}
\end{figure}

The Grover circuit is another typical periodic circuit.
In this section, we consider a 5-qubit Grover circuit
consisting of four target qubits and one auxiliary qubit.
Fig.~\ref{fig:grover_operator} shows the circuit structure of a single Grover operator. When mapped to the FakeLima noise simulator, one Grover operator contains 72 CNOT gates.

The target observable is given by
\begin{align}
    O=\ket{0000}\bra{0000}\otimes \operatorname{I}_{2}.
\end{align}

Since there is only one circuit per depth, we perform a
first-order expansion of the constant $e^{c_0}$
in Eq.~\eqref{eq93}, introducing an additional
constant to reduce the error, which gives the approximate model:
\begin{align}
    \langle P_{\beta} \rangle(k) = (a+bk)e^{c_{2}k^{2} + c_{1}k} + c. \label{eq98}
\end{align}

\begin{figure}[t]
    \centering
    \captionsetup{justification=raggedright, singlelinecheck=false}

    % (a)
    \begin{minipage}{0.48\linewidth}
        \centering
        \refstepcounter{subfig}\label{fig:device_lima}
        \begin{tikzpicture}
            \node[inner sep=0] (img) {\includegraphics[width=\linewidth]{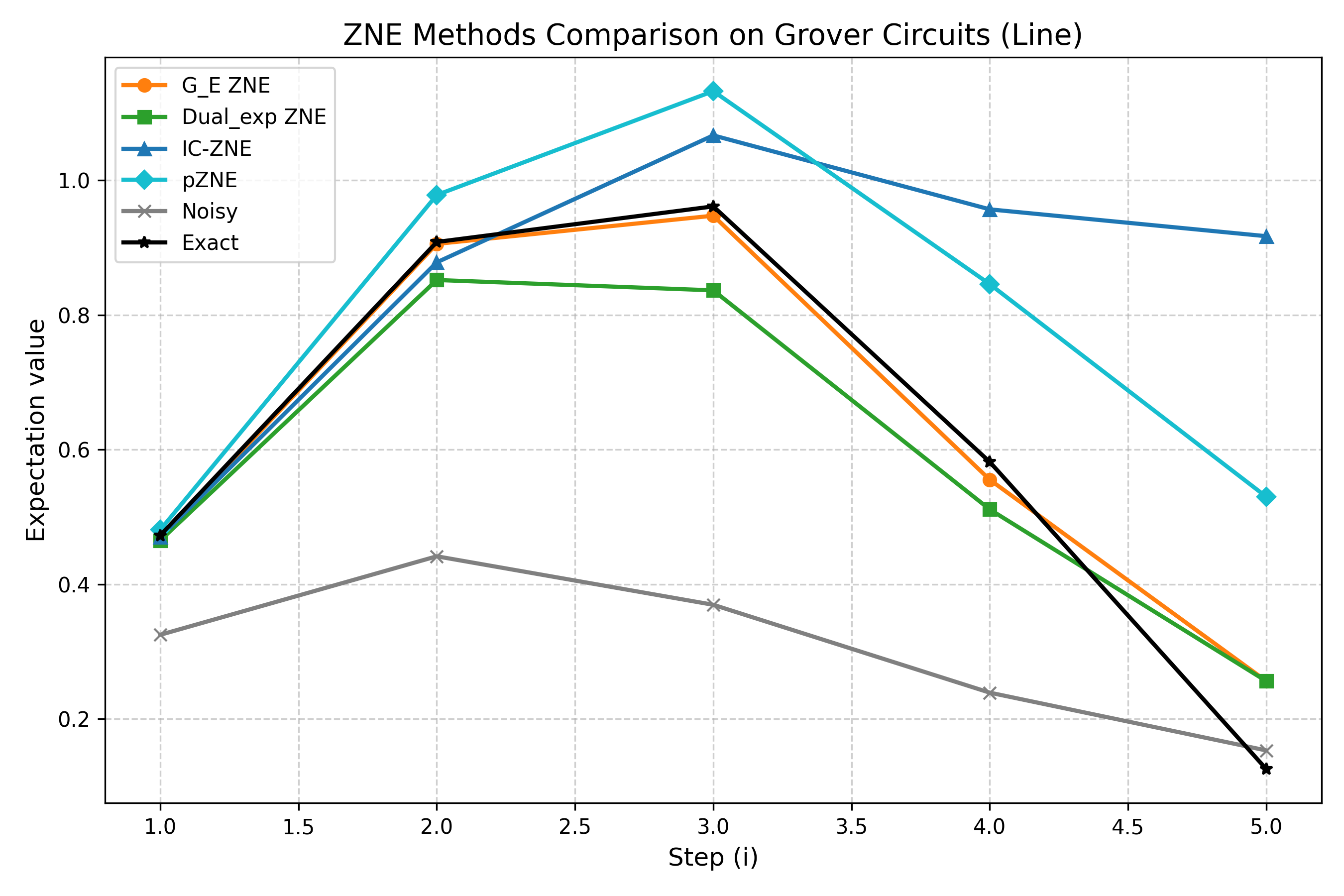}};
            \node[anchor=north west, font=\bfseries, inner sep=2pt, xshift=-6pt, yshift=-2pt] at (img.north west) {(a)};
        \end{tikzpicture}
    \end{minipage}
    \hspace{1em}
    % (b)
    \begin{minipage}{0.48\linewidth}
        \centering
        \refstepcounter{subfig}\label{fig:device_quito}
        \begin{tikzpicture}
            \node[inner sep=0] (img) {\includegraphics[width=\linewidth]{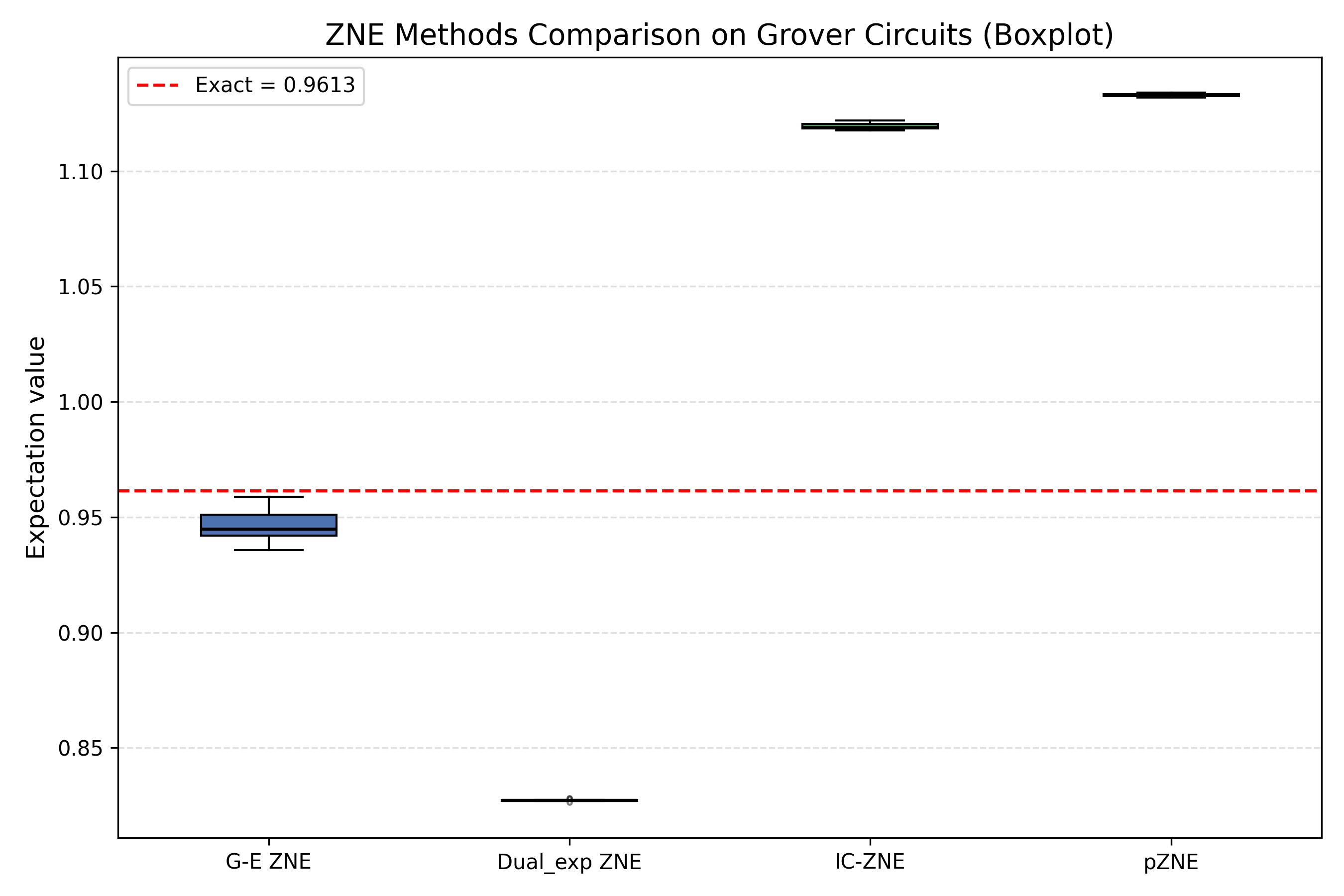}};
            \node[anchor=north west, font=\bfseries, inner sep=2pt, xshift=-6pt, yshift=-2pt] at (img.north west) {(b)};
        \end{tikzpicture}
    \end{minipage}

    \caption{Accuracy of different ZNE methods in Grover circuits.
    (a). Expectation values versus circuit iterations
    for different ZNE methods.
    (b). Box plots of expectation values for different ZNE methods
    at the first search peak.}
    \label{fig:Grover}
\end{figure}

Fig.~\ref{fig:Grover}(a) displays the expectation values for
up to five iterations, where $step=3$  corresponds to
the first search peak. Within four steps, our extrapolation
outperforms others in approaching the ideal values. At
$step=5$, all extrapolation approaches fail, likely because
the circuit depth is excessive and noise completely
alters the output structure. Fig.~\ref{fig:Grover}(b) focuses on
sampling experiments at the first search peak to assess
the stability of our model. We performed $1$\text{e6} shots
per circuit to obtain extrapolated values
and repeated
the entire procedure ten times. To mitigate readout errors,

we adopted the matrix inversion procedure \cite{Nachman:2019zfw}.
Although our method has larger variance than
the competing approaches, its worst-case result still
exceeds the average performance of the others.

\begin{figure}[t]
\centering
\includegraphics[width=0.6\textwidth]{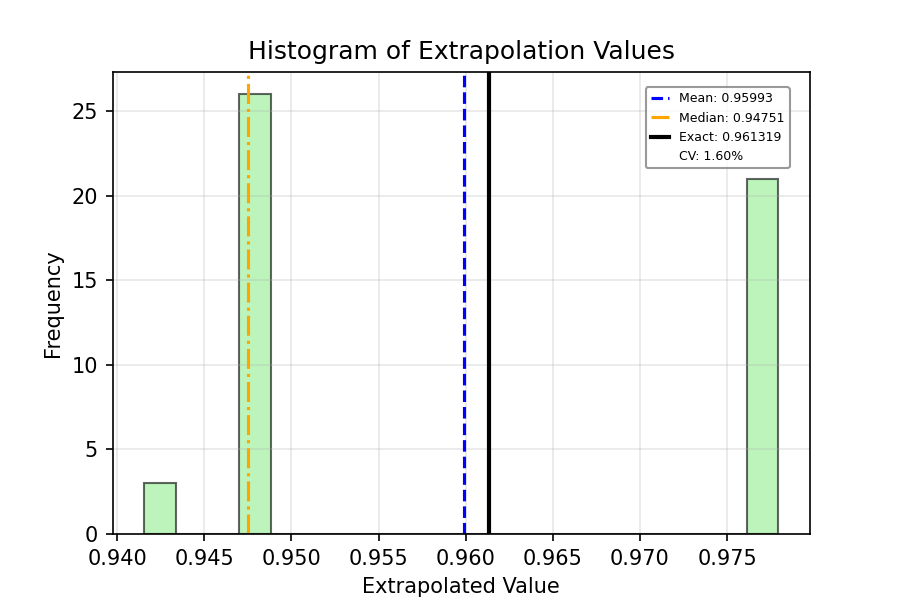}
\caption{\label{fig:stable} Histogram of extrapolated values obtained from 50 different initial parameter sets using the Levenberg--Marquardt algorithm. The mean (blue dashed), median (orange dash-dotted), exact value (black solid line), and the coefficient of variation (CV) are displayed in the legend.}
\end{figure}

Fig.~\ref{fig:stable} shows the distribution of extrapolated values obtained from 50 random initial parameter sets at $step = 3$, repeated across 10 independent sampling trials (500 fits in total). Of these, 485 fits (97\%) converge to physically reasonable values, while 15 fits (3\%) converge to non-physical local minima and are excluded from the coefficient of variation (CV) calculation. Two distinct peak regions are observed in the histogram, corresponding to two local fitting optima achieved under different initial guesses. Despite this bimodality, the CV of the well-converged fits is only 1.6\%, which is very small, indicating that the extrapolation results are stable. This demonstrates that our extrapolation model is not sensitive to the choice of initial values.


\section{Conclusions and Discussions}
This work introduces a hybrid Gaussian-exponential extrapolation model for zero-noise extrapolation in periodic quantum circuits, a structure common in quantum algorithms, including Trotterized quantum simulation, VQE parameterized circuits, QAOA, and Grover's algorithm. Starting from a Pauli path representation of noise propagation, we proved that in the asymptotic limit of large periods, the noise amplification factor converges in distribution to a log-normal law under Pauli noise channels. This statistical insight provides a rigorous foundation for augmenting the conventional exponential model with Gaussian variance corrections, yielding a more accurate description of error scaling in deep structured circuits.

We benchmarked the proposed model against standard ZNE, inverted-circuit ZNE, and purity-assisted ZNE using three representative circuit classes: Trotterized dynamics of the one-dimensional transverse-field Ising model, random circuits, and Grover circuits. The simulations, performed on Qiskit noise simulators (FakeQuito and FakeLima), show that for moderate to large circuit depths, the hybrid model gives consistently lower bias than the competing approaches. The advantage becomes more pronounced as the number of periods increases, matching the asymptotic log-normal behavior derived in the theoretical analysis.

The experimental benchmarks are restricted to 4--5 qubits and simulated noise models (FakeQuito and FakeLima), limited by the exponential cost of classical density-matrix simulation. Testing the hybrid model on larger systems and on real quantum hardware remains for future work. Second, the derivation assumes exact Pauli noise after twirling; residual coherent errors may introduce corrections not captured by our model. While the derivation assumes circuit periodicity, the experimental results indicate that the method remains effective even for modest period counts, suggesting broader applicability to circuits with approximate or emergent periodic structure. Several directions warrant further investigation. First, extending the statistical framework to non-periodic circuits may enable more general noise models. Second, the combination of this extrapolation scheme with other error mitigation techniques, such as probabilistic error cancellation and virtual distillation\cite{Huggins_2021}, is an open and promising direction. Finally, experimental validation on real quantum hardware, where non-Pauli noise components and measurement errors are present, would further establish the practical utility of the method.

The hybrid Gaussian-exponential model provides a theoretically grounded strategy for error mitigation in periodic quantum algorithms, adding to the set of available tools for reliable computation on near-term quantum devices.

\section*{Data and code availability}
The code and data that support the findings of this study are openly available at \url{https://github.com/Tao-Wa/Hybrid_Gaussian_exponential_zero_noise_extrapolation}.

\section*{Acknowledgments}
This work is supported by the National Key R\&D Program of China (Grant No. 2023YFA1009403), the National Natural Science Foundation special project of China (Grant No. 12341103) and the National Natural Science Foundation of China (Grant No. 62372444).

\bibliographystyle{unsrt}
\bibliography{apssamp}

\appendix
\section{DEVICE PROPERTIES}
\begin{figure}[ht]
    \centering
    \captionsetup{justification=raggedright, singlelinecheck=false}
    \begin{minipage}{0.48\linewidth}
        \centering
        \refstepcounter{subfig}\label{fig:Lima}
        \begin{tikzpicture}
            \node[inner sep=0] (img) {\includegraphics[width=\linewidth]{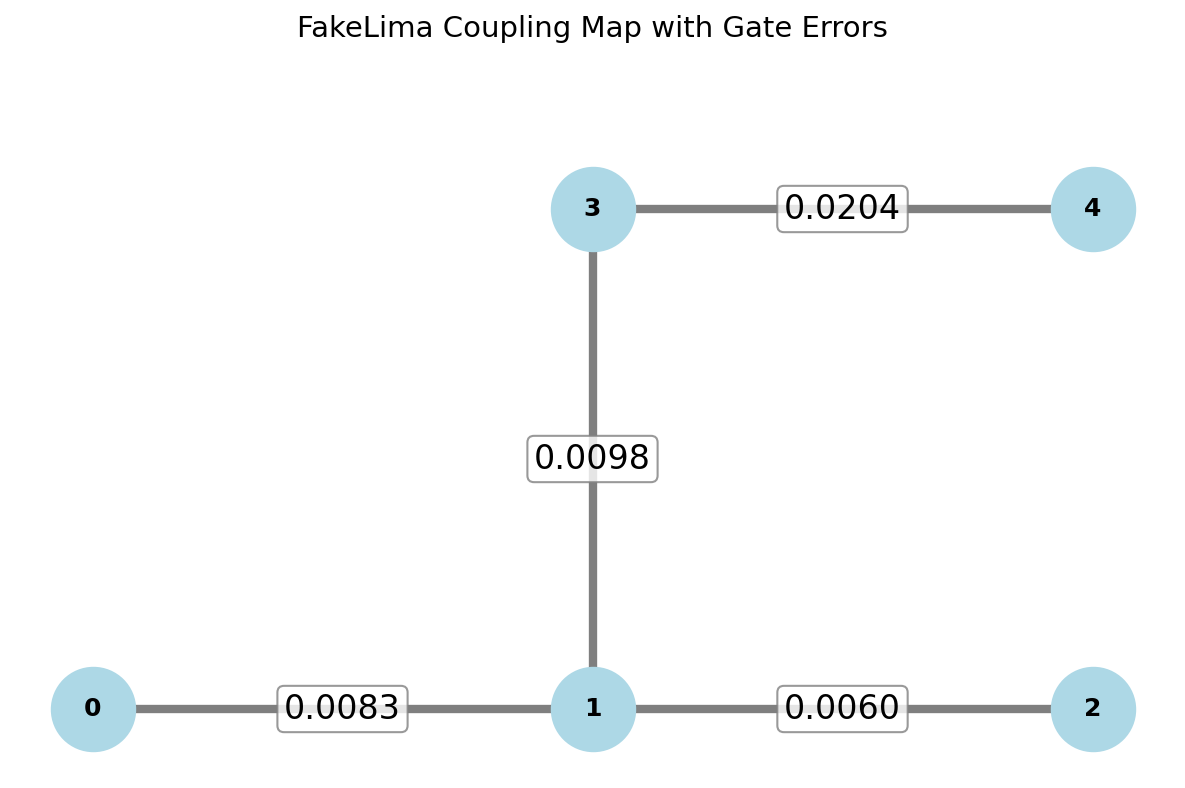}};
            \node[anchor=north west, font=\bfseries, inner sep=2pt, xshift=2pt, yshift=-2pt] at (img.north west) {(a)};
        \end{tikzpicture}
        \label{fig:device_lima}
    \end{minipage}
    \hspace{1em}
    \begin{minipage}{0.48\linewidth}
        \centering
        \refstepcounter{subfig}\label{fig:Quito}
        \begin{tikzpicture}
            \node[inner sep=0] (img) {\includegraphics[width=\linewidth]{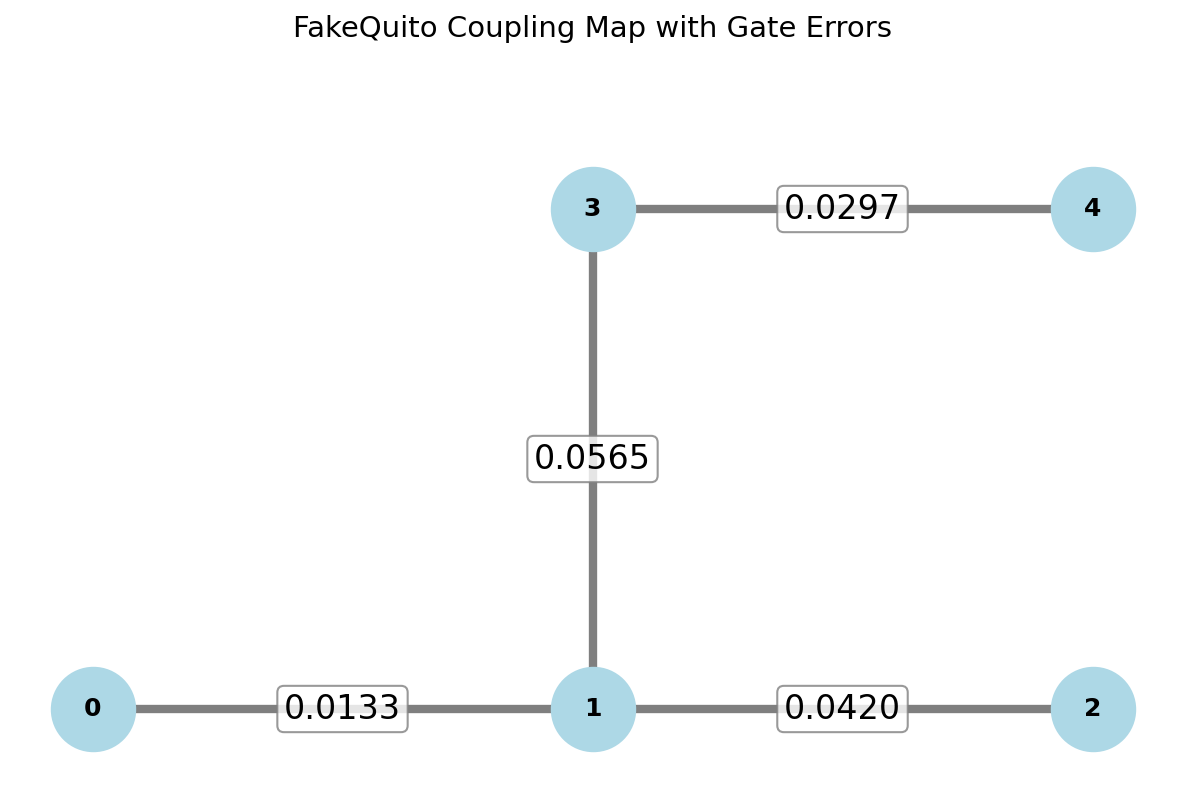}};
            \node[anchor=north west, font=\bfseries, inner sep=2pt, xshift=2pt, yshift=-2pt] at (img.north west) {(b)};
        \end{tikzpicture}
        \label{fig:device_quito}
    \end{minipage}
    \caption{Qubit connectivity and two-qubit gate error rates of the simulated IBM Quantum devices FakeLima (a) and FakeQuito (b). Each node represents a physical qubit, labeled by its index. Edges indicate the available coupling between qubits, and the numbers on each edge denote the corresponding two-qubit gate error rate extracted from the backend calibration data.}
    \label{fig:Device}
\end{figure}

Fig.~\ref{fig:Device} shows the architecture and CNOT error rates of the FakeQuito and FakeLima simulators, obtained from the Qiskit fake provider backends. Table~\ref{tab:both_devices} provides a detailed illustration of the properties of individual qubits.

\begin{table}[ht]
\centering
\captionsetup{justification=raggedright, singlelinecheck=false}
\caption{Detailed qubit parameters of the simulated device FakeLima (a) and FakeQuito (b). Anharmonicity $\eta$, relaxation time $T_1$, coherence time $T_2$, and readout fidelities $F_0$, $F_1$ are taken from the Qiskit fake backend calibration.}
\label{tab:both_devices}
\setlength{\tabcolsep}{6pt}
% (a) FakeQuito
\begin{minipage}{\textwidth}
\centering
\subcaption{}\label{tab:fakequito}
\large
\begin{tabular}{cccccccc}
\toprule
Qubit & $f_{10}$ (GHz) & $\eta$ (MHz) & $T_1$ ($\mu$s) & $T_2$ ($\mu$s) & $F_0$ & $F_1$ \\
\midrule
Q0 & 5.030 & -335.74 & 59.70 & 93.56 & 0.9882 & 0.9596 \\
Q1 & 5.128 & -318.35 & 83.06 & 115.53 & 0.9888 & 0.9712 \\
Q2 & 5.247 & -333.60 & 103.78 & 94.77 & 0.9928 & 0.9740 \\
Q3 & 5.303 & -331.24 & 43.58 & 46.46 & 0.9752 & 0.9218 \\
Q4 & 5.092 & -334.47 & 17.54 & 16.44 & 0.9808 & 0.9042 \\
\bottomrule
\end{tabular}
\end{minipage}

\vspace{0.5cm}

% (b) FakeLima
\begin{minipage}{\textwidth}
\centering
\subcaption{}\label{tab:fakelima}
\large
\begin{tabular}{cccccccc}
\toprule
Qubit & $f_{10}$ (GHz) & $\eta$ (MHz) & $T_1$ ($\mu$s) & $T_2$ ($\mu$s) & $F_0$ & $F_1$ \\
\midrule
Q0 & 5.241 & -339.47 & 98.32 & 122.41 & 0.9790 & 0.9324 \\
Q1 & 5.018 & -318.52 & 112.56 & 135.89 & 0.9956 & 0.9604 \\
Q2 & 5.183 & -331.24 & 76.45 & 88.32 & 0.9746 & 0.8032 \\
Q3 & 5.079 & -320.19 & 101.23 & 110.74 & 0.9892 & 0.9208 \\
Q4 & 4.996 & -315.08 & 128.67 & 142.18 & 0.9878 & 0.9502 \\
\bottomrule
\end{tabular}
\end{minipage}
\end{table}

\end{document}